\documentclass[twocolumn,journal]{IEEEtran}
\usepackage{ifpdf}
\usepackage{cite}
\usepackage{bbm}
\usepackage{url}
\usepackage{verbatim}
\usepackage{textcomp}
\usepackage{bm}
\ifCLASSINFOpdf
\usepackage{graphicx}
\usepackage[caption=false,font=footnotesize]{subfig}
\usepackage{epstopdf}
\else
\fi
\usepackage{float}
\usepackage{amsmath}
\usepackage{amsfonts}
\usepackage{amssymb}
\usepackage{amsthm}
\usepackage{mathtools}
\DeclareMathOperator*{\argmin}{arg\,min}
\allowdisplaybreaks
\usepackage{array}
\usepackage{stfloats}
\usepackage{placeins}
\usepackage{makecell}

\newtheorem{remark}{Remark}

\theoremstyle{definition}

\usepackage{booktabs}
\usepackage{algorithm}
\usepackage{tabularx}
\usepackage[noend]{algpseudocode}
\algrenewcommand\algorithmicrequire{\textbf{Input:}}
\algrenewcommand\algorithmicensure{\textbf{Output:}}
\algrenewcommand\algorithmiccomment[1]{\hfill// #1}

\usepackage[table,xcdraw]{xcolor}
\usepackage[shortlabels]{enumitem}
\usepackage{comment}
\usepackage{multirow}

\def\BibTeX{{\rm B\kern-.05em{\sc i\kern-.025em b}\kern-.08em
    T\kern-.1667em\lower.7ex\hbox{E}\kern-.125emX}}

\begin{document}

\title{
Extended Universal Joint Source-Channel Coding for Digital Semantic Communications: Improving Channel-Adaptability
}

\author{Eunsoo~Kim,~\IEEEmembership{Graduate Student Member,~IEEE,} Yoon~Huh,~\IEEEmembership{Graduate Student Member,~IEEE,} and Wan~Choi,~\IEEEmembership{Fellow,~IEEE}
\thanks{E. Kim, Y. Huh and W. Choi are with the Department of Electrical and Computer Engineering, Seoul National University (SNU), and the Institute of New Media and Communications, SNU, Seoul 08826, Korea (e-mail: {es3783, mnihy621, wanchoi}@snu.ac.kr) \emph{(Corresponding author: Wan Choi) }} }

\maketitle

\begin{abstract}
Deep learning (DL)-based joint source--channel coding (JSCC), particularly vector quantization (VQ)-based JSCC, enables efficient semantic communication by mapping high-dimensional feature vectors into compact codeword indices for digital modulation. However, existing methods, including universal JSCC (uJSCC), rely on fixed, modulation-specific encoders, decoders, and codebooks, limiting adaptation to fine-grained channel variations. We propose an extended universal JSCC (euJSCC) framework that enables adaptive transmission across both modulation orders and fine-grained channel conditions within a single model using hypernetwork-based normalization and a dynamic codebook generation (DCG) network that refines modulation-specific base codebooks according to channel quality indicator (CQI) feedback. To support time-varying channels with periodic CQI feedback, euJSCC employs an inner--outer encoder--decoder architecture, where the outer modules capture long-term channel statistics and the inner modules refine feature vectors to align with block-wise codebooks. A two-phase training strategy first pretrains the outer modules and DCG network over additive white Gaussian noise (AWGN) channels to establish stable feature--codebook alignment and then fine-tunes the full model over time-varying channels. Experiments demonstrate that euJSCC consistently outperforms state-of-the-art channel-adaptive digital JSCC schemes in image transmission, while additional evaluations under practical channel and feedback conditions and on a downstream task further validate its effectiveness.
\end{abstract}


\IEEEpeerreviewmaketitle

\section{Introduction}\label{sec:Introduction}

The sixth-generation (6G) vision targets ultra-dense connectivity, near-zero latency, and high reliability, but conventional bit-level transmission is inefficient for complex task-oriented data in dynamic, bandwidth-limited environments. Semantic communication~\cite{shao2024theory, shi2023task, seo2025contextual, huh2025feature, huh2026differential} therefore prioritizes task-relevant information over bit-perfect reconstruction, and deep learning (DL)-based joint source–channel coding (JSCC) enables end-to-end encoder–decoder design by mapping semantic feature vectors directly to channel symbols and performing target tasks from channel-impaired feature vectors, which removes separate source and channel codecs while improving robustness and spectral efficiency~\cite{farsad2018deep}.

However, transmitting continuous-valued feature vectors, referred to as analog JSCC, requires analog modulation that is incompatible with modern digital infrastructures, and channel noise directly distorts feature values, potentially causing unstable task performance under poor channels or multi-hop networks~\cite{zhang2025toward}. These drawbacks motivate digital JSCC, where feature vectors are quantized into bitstreams and transmitted via standard digital modulation (e.g., BPSK, 4-QAM, 16-QAM).

Existing digital JSCC methods are generally categorized into scalar quantization (SQ) and vector quantization (VQ) approaches. SQ-based digital JSCC (SQ-JSCC) discretizes feature vectors element-wise, mapping each scalar element to a learned quantization level~\cite{guo2024digital} or a point in a modulation-specific constellation~\cite{tung2022deepjscc, zhang2024analog}. In contrast, VQ-based JSCC (VQ-JSCC) jointly quantizes multi-dimensional feature vectors using learned codebooks, effectively capturing inter-dimensional correlations and achieving superior rate–distortion performance~\cite{xie2023robust, zheng2025deep, huh2025universal, huh2026markov}.

In practice, wireless channels are inherently time-varying due to fading, interference, and user mobility. Training and maintaining separate JSCC models for every possible channel condition would incur prohibitive memory and computational costs. Therefore, a unified JSCC model capable of adapting to diverse channel conditions is essential for practical deployment. However, existing studies~\cite{tung2022deepjscc, xie2023robust} lack such adaptability, leading to severe performance degradation under unseen channel environments.

To address channel variations in SQ-JSCC, \cite{guo2024digital} proposed a framework in which the network split point is adaptively selected according to the channel conditions, and the intermediate feature vectors at the chosen depth are scalar-quantized and transmitted. Building on neural network split inference that considers transmitter-side computational constraints and signal-to-noise ratio (SNR) conditions as in~\cite{lee2023wireless}, the authors in \cite{guo2024digital} further posited that feature vectors produced at different network depths contain different semantic information and exhibit varying levels of noise resistance. Accordingly, a learned policy network determines the split point according to the SNR, while the quantization levels are individually optimized for each split point to account for the resulting differences in feature distributions. However, the encoder--decoder process itself does not adapt to the channel condition, thereby limiting overall adaptability.

Following the channel adaptation principle of modern digital communication systems, another line of research has investigated JSCC schemes that adaptively support multiple modulation orders according to channel conditions. \cite{zhang2024analog} introduced a multi-order digital joint coding-modulation (MDJCM) built upon nonlinear transform coding (NTC)~\cite{balle2020nonlinear}, which adjusts the length of feature vectors according to their estimated entropy. It employs modulation- and SNR-conditioned modules based on squeeze-and-excitation (SE) layers~\cite{hu2018squeeze}, which adaptively reweight feature vector channels according to the given transmission conditions. However, MDJCM quantizes 2-dimensional segments of each feature vector using the modulation constellation as a fixed 2-dimensional codebook, as in~\cite{tung2022deepjscc}. Although higher modulation orders reduce quantization error through finer constellations, the feature vector dimensionality does not increase with modulation orders, limiting the amount of information that can be represented. As a result, the task performance gain saturates. Moreover, since the length of each feature vector varies according to its estimated entropy, side information indicating the feature vector lengths must be reliably conveyed to the receiver. This requires an additional reliable transmission mechanism for the side information, thereby increasing both system complexity and communication overhead.

On the other hand, since each \textit{multi-dimensional} feature vector is mapped to a single codeword index in VQ-JSCC, even one bit error can corrupt the transmitted index, leading to the receiver decoding a vastly different codeword. This high sensitivity severely distorts the alignment between the extracted feature vector and the received codeword, causing substantial task performance degradation. Consequently, VQ-JSCC is inherently vulnerable to channel impairments, particularly under low-SNR or fading conditions. To enhance robustness, it is therefore essential to adapt not only the encoder–decoder process but also the codebook itself, ensuring that the alignment between feature vectors and codewords remains consistent across varying channel conditions. In other words, VQ-JSCC must be capable of channel-adaptively aligning the feature vector space with the multi-dimensional codebook to achieve optimal task performance.

To address this challenge, a feedback-driven codebook optimization (FDCO)~\cite{zheng2025deep} proposed adapting codewords using an SNR-conditioned multilayer perceptron (MLP) trained after pretraining the encoder, decoder, and codebook. However, this method assumes a fixed modulation scheme and independently updates each codeword via a shared MLP while freezing the encoder and decoder during adaptation. As a result, the joint alignment between features and codewords is disrupted, ultimately limiting the achievable performance of VQ-JSCC under diverse channel conditions.

Another work~\cite{huh2025universal} introduced a universal JSCC (uJSCC) framework that extends VQ-JSCC to support multiple modulation orders with negligible parameter overhead. In uJSCC, modulation-specific codebooks are paired with switchable normalization layers in the encoder and decoder. Since distinct codebooks yield different feature vector statistics, uJSCC employs a dedicated normalization layer for each modulation to stabilize these statistics and ensure adaptability across modulation conditions. However, a key limitation is that the encoder, decoder, and codebooks remain fixed for a given modulation order, which is used within its predefined SNR range. In practice, even when the same modulation is used, different SNR values produce different error characteristics. Thus, leveraging fine-grained channel conditions within each modulation region can further improve the adaptability of the JSCC framework.

To address the limited fine-grained channel adaptability, we therefore propose an \textit{extended universal JSCC (euJSCC)}, which enhances the channel adaptability of uJSCC by jointly adapting the encoder--decoder feature vector representation and VQ codebook to both modulation orders and fine-grained channel conditions. Specifically, the switchable normalization layer is extended from modulation-conditioned operation to operation conditioned on both the modulation order and fine-grained channel information, enabling fine-grained normalization during encoding/decoding. Moreover, euJSCC employs dynamic codebook generation that refines modulation-specific base codebooks according to fine-grained channel information, enabling fine-grained channel-adaptive VQ.

Furthermore, contrary to the assumption of static channels with constant fading and SNR per source transmission, we consider time-varying fading channels. To enable fine-grained adaptation under such channels, we consider periodic channel information feedback from the receiver to the transmitter. euJSCC then adapts using the most recently available channel information and updates its adaptation upon each feedback. However, performing such fine-grained adaptation throughout the entire JSCC encoder--decoder at every feedback would incur substantial computational overhead. Inspired by the coarse-to-fine refinement principle in~\cite{li2025coarse}, euJSCC extends coarse-to-fine adaptation to VQ-JSCC via an inner--outer encoder--decoder architecture: the outer encoder--decoder adapts to the long-term channel statistics and a coarse modulation order, while the inner encoder--decoder refines the feature vector channel dimensions and statistics according to the short-term channel condition and corresponding modulation order. Since such short-term adaptation is performed throughout transmission, all adaptive components are designed to be parameter- and computation-efficient to support frequent adaptation. Finally, we comprehensively evaluate the proposed framework under time-varying channels while accounting for practical communication scenarios, including feedback delay and imperfect channel information.
\vspace{-3mm}
\subsection{Contributions}
The contributions of this article are summarized as follows:
\begin{itemize}

\item \textbf{Hypernetwork-based channel-adaptive normalization:} euJSCC enables unified adaptation to modulation orders and fine-grained channel conditions through a parameter-efficient hypernetwork-based normalization layer that generates modulation- and channel-dependent parameters with minimal additional overhead.

\item \textbf{Dynamic channel-adaptive VQ codebook generation:} A dynamic VQ codebook generation network adapts modulation-specific base codebooks to fine-grained channel conditions through intra- and inter-codeword refinement rather than full codebook regeneration.

\item \textbf{Efficient adaptation over time-varying channels:} euJSCC employs an inner--outer encoder--decoder structure that separates long- and short-term channel adaptation under periodic channel information feedback. A two-phase training strategy first establishes stable feature--codebook alignment over AWGN channels without the inner encoder--decoder and then finetunes the full model over time-varying channels.

\item \textbf{Comprehensive experiments:} Extensive experiments demonstrate that euJSCC consistently outperforms existing channel-adaptive digital JSCC schemes under time-varying channels while maintaining parameter and computational efficiency, with further evaluations under practical channel and feedback conditions and on a downstream semantic task.

\end{itemize}

\vspace{-3mm}
\section{System Model}\label{sec:System Model}

\begin{figure}[t]
    \centering
    \includegraphics[width=0.9\linewidth]{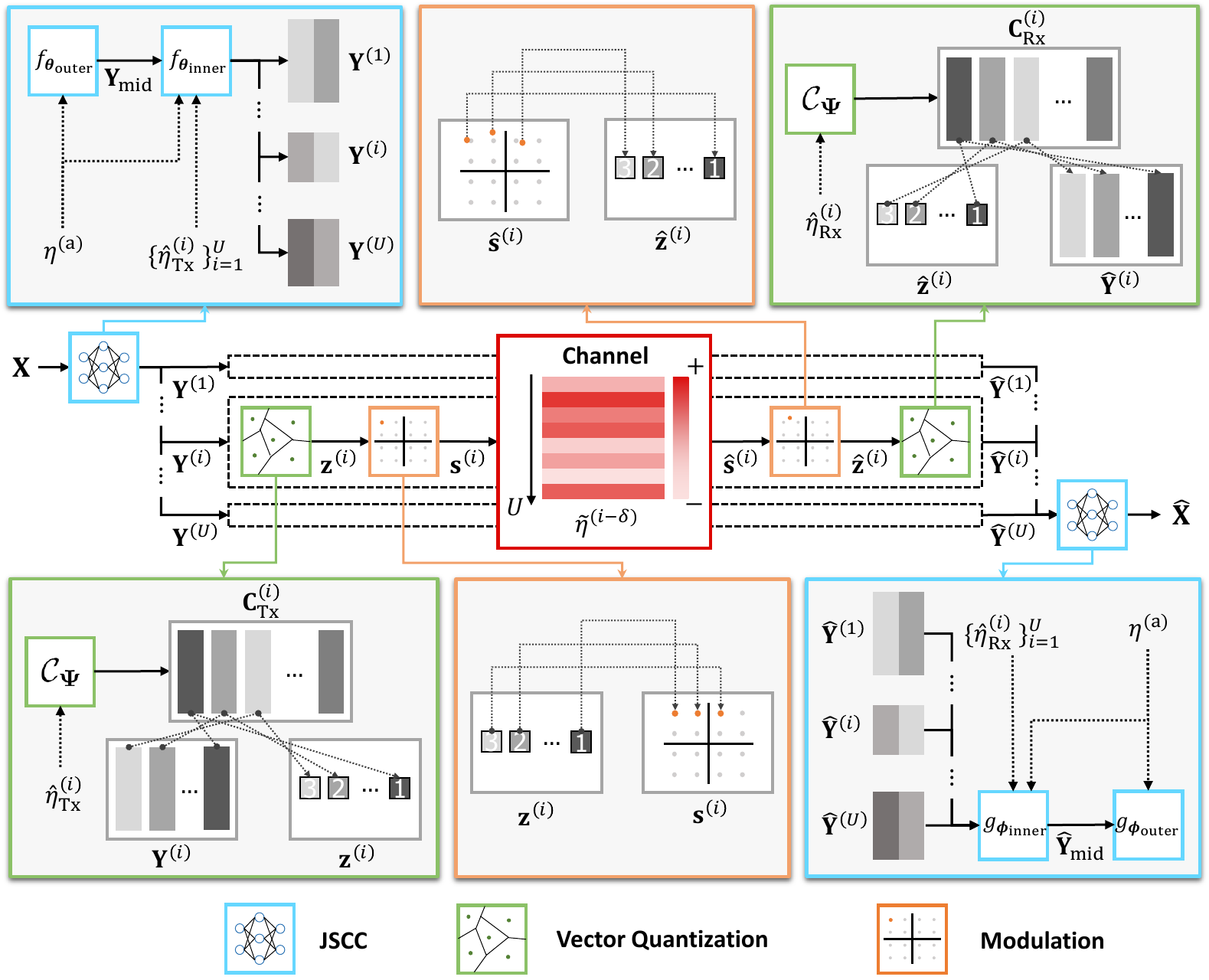}
    \vspace{-1mm}
    \caption{System overview of euJSCC, which performs transmission block-wise feature vector encoding-decoding, VQ, and modulation.}
    \label{fig:system model}
    \vspace{-5mm}
\end{figure}

\subsection{System Preliminaries and Channel Model}
\label{subsec:System Preliminaries and Channel Model}

We consider a point-to-point digital semantic communication system for image transmission over time-varying flat fading channels. The system comprises a JSCC encoder--decoder, a vector quantizer--dequantizer, a VQ codebook generator, and a modulator--demodulator pair. To enable adaptation to time-varying channel conditions, updating the transmission strategy for every transmitted symbol would incur excessive feedback and adaptation overhead. Accordingly, the receiver periodically feeds back a channel quality indicator (CQI), and the transmitter updates its adaptation whenever new CQI becomes available while using the most recently available CQI between feedbacks. In practical wireless systems, pilot symbols serving as reference signals are periodically transmitted to estimate the channel condition, and the resulting channel quality is reported through CQI for channel adaptation~\cite{3gpp.38.214}. Consistent with this procedure, in our system, one pilot symbol is transmitted for every $T_p$ data symbols. We define a \textit{transmission block} as comprising $T$ data symbols and the additional pilot symbols, with CQI feedback and channel adaptation updated once per block. The value of $T$ can be selected considering the channel dynamics and feedback constraints, e.g., a smaller $T$ may be used under rapidly varying channels to enable more frequent CQI feedback and adaptation at the cost of increased feedback overhead. Let $N$ denote the total number of transmitted data symbols. The transmission is divided into $U=\lceil N/T\rceil$ transmission blocks indexed by $i\in[1{:}U]$. For notational simplicity, we first assume that $N$ is an integer multiple of $T$, so that all $U$ transmission blocks have an equal length of $T$ data symbols. The general case where $N$ is not an integer multiple of $T$ is described in Section~\ref{subsec:Implementation of the Encoder-Decoder Architecture}.

For the $i$-th transmission block, let $h_{t}^{(i)}$ denote the channel coefficient for the $t$-th data symbol, where $t\in[1{:}T]$. The instantaneous SNR of the $t$-th data symbol is given by $\frac{|h_{t}^{(i)}|^{2}P}{\sigma^{2}}$, where $P$ denotes the average transmit power per modulated symbol and $\sigma^{2}$ denotes the noise variance of the considered time-varying channel. Using the received pilot symbols, the receiver first estimates the channel at the pilot symbols via linear minimum mean-square-error (LMMSE) estimation and then interpolates the estimates for the intervening data symbols, yielding $\{\hat{h}_{t}^{(i)}\}_{t=1}^{T}$.

For CQI computation, the receiver summarizes the channel quality over transmission block $i$ using the block-wise SNR $\eta^{(i)}=10\log_{10}\left(\frac{P}{T\sigma^{2}}\sum_{t=1}^{T}|\hat{h}_{t}^{(i)}|^{2}\right)$ (dB), obtained by averaging the estimated instantaneous SNRs over its $T$ data symbols. The receiver quantizes the measured block-wise SNR $\eta^{(i)}$ into the CQI $\tilde{\eta}^{(i)}=\mathcal{Q}_{\mathsf{CQI}}\big(\eta^{(i)}\big)$ using $B_{\mathsf{CQI}}$ feedback bits, where $\mathcal{Q}_{\mathsf{CQI}}(\cdot)$ denotes a CQI quantization function with $2^{B_{\mathsf{CQI}}}$ quantization levels\footnote{For simplicity, we employ uniform quantization, where the quantization range is selected to cover $95\%$ of the block-wise SNR realizations under the considered system setting, and $B_{\mathsf{CQI}}$ is set to $4$ following the 4-bit CQI representation specified in~\cite{3gpp.38.214}.}. The receiver feeds back the quantized CQI $\tilde{\eta}^{(i)}$ once per transmission block.

Due to feedback delay, $\tilde{\eta}^{(i)}$ may not be immediately available at the transmitter. We express the feedback delay in units of transmission blocks by $\delta\in\mathbb{Z}_{\geq 0}$, where $\delta=0$ represents the case without feedback delay. For example, $\delta=1$ corresponds to using the CQI obtained from the immediately preceding transmission block. For $i-\delta\leq0$, the required $\tilde{\eta}^{(i-\delta)}$ values are assumed to have been obtained from channel observations and fed back prior to the current transmission. Furthermore, since the feedback delay $\delta$ is known to both the transmitter and receiver as specified in~\cite{3gpp.38.214}, the receiver uses the CQI generated $\delta$ blocks earlier for its channel-adaptive operations. Thus, the available CQIs for the $i$-th transmission block are given by
\begin{equation}
    \hat{\eta}_{\star}^{(i)} =
    \begin{cases}
        \mathcal{W}\big(\tilde{\eta}^{(i-\delta)}\big), & \star=\mathsf{Tx},\\
        \tilde{\eta}^{(i-\delta)}, & \star=\mathsf{Rx},
    \end{cases}
    \qquad
    i\in[1{:}U],
    \label{eq:cqi_feedback_delay}
\end{equation}
where $\hat{\eta}_{\mathsf{Tx}}^{(i)}$ and $\hat{\eta}_{\mathsf{Rx}}^{(i)}$ denote the CQIs available at the transmitter and receiver, respectively, and $\mathcal{W}(\cdot)$ represents CQI transmission over the control channel\footnote{In practical wireless systems, CQI is conveyed through a control channel and protected by channel coding to ensure highly reliable control signaling. Accordingly, the 4-bit CQI in our system model is transmitted with the (32,4) short-block channel coding and BPSK specified in~\cite{3gpp.38.212}.}. The transmitter- and receiver-side CQIs are identical unless a feedback error occurs. We assume that the long-term channel statistics $\mathbb{E}[|h_{t}^{(i)}|^{2}] = 1$ are known to both the transmitter and the receiver, and thus the average SNR, defined as $\eta^{(\mathsf{a})}=10\log_{10}\left(\frac{\mathbb{E}[|h_{t}^{(i)}|^{2}]P}{\sigma^{2}}\right)$ (dB), is known at both sides.

The system adaptively switches among $K$ modulation orders $\{m_{1},m_{2},\dots,m_{K}\}$, where $m_{k}$ represents the $k$-th modulation order and, without loss of generality, $m_{1}<m_{2}<\cdots<m_{K}$. The modulation order for the $i$-th transmission block, denoted by $m_{k^{(i)}}$, is selected according to the transmitter-side CQI $\hat{\eta}_{\mathsf{Tx}}^{(i)}$ and a predefined set of SNR boundaries $\mathbf{b}=[b_{1},\dots,b_{K-1}]$\footnote{Following the practice of conventional adaptive modulation and coding (AMC) strategy, we select the modulation order using a predefined ordered boundary set $\mathbf{b}$. The proposed framework is not tied to particular boundary values, and the boundary values used in our experiments as well as the impact of different boundary settings are presented in Section~\ref{subsec:Performance Under Various Channel and System Conditions}.}:
\begin{align}
    \label{eq:modulation-order-selection}
    m_{k^{(i)}}=
    \begin{cases}
        m_{1}, & \text{if } \hat{\eta}_{\mathsf{Tx}}^{(i)}<b_{1},\\
        m_{k}, & \text{if } b_{k-1}\leq\hat{\eta}_{\mathsf{Tx}}^{(i)}<b_{k},\\
        m_{K}, & \text{if } \hat{\eta}_{\mathsf{Tx}}^{(i)}\geq b_{K-1},
    \end{cases}
\end{align}
where $k\in[2{:}K-1]$ and $k^{(i)}$ denotes the modulation-order index for the $i$-th block. As in 3GPP NR~\cite{3gpp.38.214}, the selected modulation order can be indicated to the receiver through the MCS information included in the control signaling, ensuring that the transmitter and receiver use the same modulation order even when a CQI feedback error occurs.

\vspace{-3mm}
\subsection{euJSCC Communication Process}
\label{subsec:euJSCC Communication Process}

With the above settings, the semantic communication proceeds as follows. The end-to-end operation of the euJSCC system is also summarized in \textbf{Algorithm~\ref{alg:euJSCC-Forward-FAST}.}

\subsubsection{Adaptive Semantic Encoding and Vector Quantization}
\label{subsubsec:Adaptive Semantic Encoding and Vector Quantization}

Given a source image $\mathbf{X}\in\mathbb{R}^{C\times H\times W}$, where $C$, $H$, and $W$ denote the number of image channels, height, and width, respectively, the JSCC encoder $\mathcal{F}_{\boldsymbol{\Theta}}$ generates block-wise adapted feature vectors sequentially over $U$ transmission blocks. For the $i$-th transmission block, the encoder generates the adapted feature vectors $\mathbf{Y}^{(i)}$ by exploiting $\eta^{(\mathsf{a})}$, $\hat{\eta}_{\mathsf{Tx}}^{(i)}$ and $k^{(i)}$. Consequently, the set of block-wise feature vectors $\{\mathbf{Y}^{(i)}\}_{i=1}^{U}$ is produced.

To effectively respond to short-term channel fading variations, the encoder is decomposed into an outer encoder $f_{\boldsymbol{\theta}_{\mathsf{outer}}}$ and an inner encoder $f_{\boldsymbol{\theta}_{\mathsf{inner}}}$, parameterized respectively by $\boldsymbol{\theta}_{\mathsf{outer}}$ and $\boldsymbol{\theta}_{\mathsf{inner}}$, so that $\boldsymbol{\Theta}=\{\boldsymbol{\theta}_{\mathsf{outer}},\boldsymbol{\theta}_{\mathsf{inner}}\}$. The outer encoder adapts to the long-term channel statistics captured by $\eta^{(\mathsf{a})}$ and produces an intermediate feature vector sequence, while the inner encoder refines this sequence in a sequential block-wise manner, producing the $i$-th block's feature vectors conditioned on $\eta^{(\mathsf{a})}$, $\hat{\eta}_{\mathsf{Tx}}^{(i)}$ and $k^{(i)}$. The implementation of the outer--inner encoder is provided in Section~\ref{subsec:Implementation of the Encoder-Decoder Architecture}.

The detailed process is as follows. The outer encoder outputs an intermediate feature vector sequence $\mathbf{Y}_{\mathsf{mid}}=[\mathbf{y}_{\mathsf{mid},1},\dots,\mathbf{y}_{\mathsf{mid},N}]^{\top}\in\mathbb{R}^{N\times D_{k^{(\mathsf{a})}}}$, conditioned on $\eta^{(\mathsf{a})}$:
\begin{equation}
    \mathbf{Y}_{\mathsf{mid}}
    =
    f_{\boldsymbol{\theta}_{\mathsf{outer}}}
    (\mathbf{X},\eta^{(\mathsf{a})}),
\end{equation}
where $k^{(\mathsf{a})}$ is the modulation index selected by applying $\eta^{(\mathsf{a})}$ to the selection rule in \eqref{eq:modulation-order-selection}, and $D_{k^{(\mathsf{a})}}\in\{D_{1},\dots,D_{K}\}$ is the corresponding feature vector channel dimension. We assume $D_{1}<\cdots<D_{K}$, indicating that higher modulation orders are assigned higher-dimensional feature vectors.

The intermediate sequence is then partitioned into $U$ non-overlapping segments of length $T$ as $\mathbf{Y}_{\mathsf{mid}}^{(i)} = [\mathbf{y}_{\mathsf{mid},(i-1)T+1},\dots,\mathbf{y}_{\mathsf{mid},iT}]^{\top} \in \mathbb{R}^{T\times D_{k^{(\mathsf{a})}}}$ for $i\in[1{:}U]$. The segments are then processed sequentially. For transmission block $i$, $f_{\boldsymbol{\theta}_{\mathsf{inner}}}$ refines $\mathbf{Y}_{\mathsf{mid}}^{(i)}$ into $\mathbf{Y}^{(i)}\in\mathbb{R}^{T\times D_{k^{(i)}}}$ based on $\eta^{(\mathsf{a})}$, $\hat{\eta}_{\mathsf{Tx}}^{(i)}$ and $k^{(i)}$:
\begin{equation}
    \mathbf{Y}^{(i)}
    =
    f_{\boldsymbol{\theta}_{\mathsf{inner}}}
    \big(
        \mathbf{Y}_{\mathsf{mid}}^{(i)},
        \eta^{(\mathsf{a})},
        \hat{\eta}_{\mathsf{Tx}}^{(i)}, k^{(i)}
    \big).
\end{equation}
Here, $D_{k^{(i)}}\in\{D_{1},\dots,D_{K}\}$ denotes the feature vector channel dimension corresponding to the modulation-order index $k^{(i)}$ selected for the $i$-th transmission block.

Next, for digital transmission, the extracted feature vectors $\mathbf{Y}^{(i)}$ are quantized through VQ. A dynamic codebook generator (DCG) network produces the block-wise codebook $\mathbf{C}_{\mathsf{Tx}}^{(i)} = [\mathbf{c}_{\mathsf{Tx},1}^{(i)},\dots,\mathbf{c}_{\mathsf{Tx},m_{k^{(i)}}}^{(i)}]^{\top} \in \mathbb{R}^{m_{k^{(i)}}\times D_{k^{(i)}}}$ corresponding to $k^{(i)}$ and $\hat{\eta}_{\mathsf{Tx}}^{(i)}$. Here, $\mathbf{c}_{\mathsf{Tx},j}^{(i)}\in\mathbb{R}^{D_{k^{(i)}}}$ denotes the $j$-th codeword of the $i$-th transmission block, and the number of codewords of $\mathbf{C}_{\mathsf{Tx}}^{(i)}$ is set equal to the modulation order $m_{k^{(i)}}$. Specifically, $\hat{\eta}_{\mathsf{Tx}}^{(i)}$ and $k^{(i)}$ are fed into the DCG network $\mathcal{C}_{\boldsymbol{\Psi}}$, parameterized by $\boldsymbol{\Psi}$, generating the codebook as
\begin{equation}
    \mathbf{C}_{\mathsf{Tx}}^{(i)}
    =
    \mathcal{C}_{\boldsymbol{\Psi}}
    \big(\hat{\eta}_{\mathsf{Tx}}^{(i)}, k^{(i)}\big).
\end{equation}
The organization of the DCG network is described in Section~\ref{subsec:Dynamic Codebook Generator}.

A quantization index of the $t$-th feature vector $\mathbf{y}_{t}^{(i)}$ is obtained as $z_{t}^{(i)} = \argmin_{j\in[1{:}m_{k^{(i)}}]}\|\mathbf{y}_{t}^{(i)}-\mathbf{c}_{\mathsf{Tx},j}^{(i)}\|_{2}^{2}$ for $t\in[1{:}T]$. The resulting index sequence of transmission block $i$ is $\mathbf{z}^{(i)} = [z_{1}^{(i)},z_{2}^{(i)},\dots,z_{T}^{(i)}]^{\top}\in[1{:}m_{k^{(i)}}]^{T}$, which is mapped one-to-one to a modulation constellation $\mathbb{S}_{m_{k^{(i)}}}=\{s_{1},\dots,s_{m_{k^{(i)}}}\}\subset\mathbb{C}$. The transmitter forms a modulated symbol sequence $\mathbf{s}^{(i)}=[s_{z_{1}^{(i)}},s_{z_{2}^{(i)}},\dots,s_{z_{T}^{(i)}}]^{\top}\in\mathbb{C}^{T}$ for transmission block $i$, normalized to satisfy the average transmit power constraint $\frac{1}{T}\mathbb{E}\left[\|\mathbf{s}^{(i)}\|_{2}^{2}\right]=P$. The transmitter then inserts a pilot symbol after every $T_p$ symbols in $\mathbf{s}^{(i)}$ and transmits the resulting sequence.

\subsubsection{Wireless Channel Input--Output Relation}
\label{subsubsec:Wireless Channel Input-Output Relation}

The wireless channel input--output relationship for the data symbols in the $i$-th transmission block is expressed as
\begin{equation}
    \label{eq:channel relation}
    \hat{\mathbf{s}}^{(i)}
    =
    \mathbf{H}^{(i)}\mathbf{s}^{(i)}
    +
    \mathbf{w}^{(i)},
\end{equation}
where $\hat{\mathbf{s}}^{(i)}=[\hat{s}_{1}^{(i)},\hat{s}_{2}^{(i)},\dots,\hat{s}_{T}^{(i)}]^{\top}\in\mathbb{C}^{T}$ denotes the received symbol sequence corresponding to $\mathbf{s}^{(i)}$, $\mathbf{H}^{(i)}=\operatorname{diag}(h_{1}^{(i)},\dots,h_{T}^{(i)})\in\mathbb{C}^{T\times T}$ denotes the channel matrix, where $\operatorname{diag}(a_{1},\dots,a_{T})$ denotes the diagonal matrix with diagonal entries $\{a_{t}\}_{t=1}^T$, and $\mathbf{w}^{(i)}\sim\mathcal{CN}(\mathbf{0},\sigma^{2}\mathbf{I}_{T})$ denotes the AWGN vector. Using the channel estimates $\{\hat{h}_{t}^{(i)}\}_{t=1}^{T}$ obtained from the received pilot symbols, the receiver forms the estimated channel matrix $\hat{\mathbf{H}}^{(i)}=\operatorname{diag}(\hat{h}_{1}^{(i)},\dots,\hat{h}_{T}^{(i)})$ and performs zero-forcing (ZF) equalization, yielding $\tilde{\mathbf{s}}^{(i)}=(\hat{\mathbf{H}}^{(i)})^{-1}\hat{\mathbf{s}}^{(i)}=(\hat{\mathbf{H}}^{(i)})^{-1}\mathbf{H}^{(i)}\mathbf{s}^{(i)}+(\hat{\mathbf{H}}^{(i)})^{-1}\mathbf{w}^{(i)}$.
Here, $\tilde{\mathbf{s}}^{(i)}=[\tilde{s}_{1}^{(i)},\dots,\tilde{s}_{T}^{(i)}]^{\top}\in\mathbb{C}^{T}$ denotes the equalized symbol sequence.

\subsubsection{Adaptive Vector Dequantization and Semantic Decoding}
\label{subsubsec:Adaptive Vector Dequantization and Semantic Decoding}

Symmetrically to the encoder, the receiver employs a JSCC decoder $\mathcal{G}_{\boldsymbol{\Phi}}$ comprising an inner decoder $g_{\boldsymbol{\phi}_{\mathsf{inner}}}$ and an outer decoder $g_{\boldsymbol{\phi}_{\mathsf{outer}}}$, parameterized respectively by $\boldsymbol{\phi}_{\mathsf{inner}}$ and $\boldsymbol{\phi}_{\mathsf{outer}}$, so that $\boldsymbol{\Phi}=\{\boldsymbol{\phi}_{\mathsf{inner}},\boldsymbol{\phi}_{\mathsf{outer}}\}$. The implementation of the inner--outer decoder is provided in Section~\ref{subsec:Implementation of the Encoder-Decoder Architecture}.

For the $i$-th transmission block, $\tilde{\mathbf{s}}^{(i)}$ is then demodulated to obtain the estimated index sequence $\hat{\mathbf{z}}^{(i)}=[\hat{z}_{1}^{(i)},\dots,\hat{z}_{T}^{(i)}]^{\top}\in[1{:}m_{k^{(i)}}]^{T}$. Using the block-wise codebook $\mathbf{C}_{\mathsf{Rx}}^{(i)}$, generated by $\mathcal{C}_{\boldsymbol{\Psi}}$ based on $\hat{\eta}_{\mathsf{Rx}}^{(i)}$ and $k^{(i)}$ at the receiver, the received feature vectors are reconstructed as $\hat{\mathbf{Y}}^{(i)}=[\mathbf{c}^{(i)}_{\mathsf{Rx},\hat{z}_{1}^{(i)}},\dots,\mathbf{c}^{(i)}_{\mathsf{Rx},\hat{z}_{T}^{(i)}}]^{\top}\in\mathbb{R}^{T\times D_{k^{(i)}}}$. The demodulated and dequantized feature vectors $\hat{\mathbf{Y}}^{(i)}\in\mathbb{R}^{T\times D_{k^{(i)}}}$ are first refined by the inner decoder $g_{\boldsymbol{\phi}_{\mathsf{inner}}}$ to reconstruct the intermediate feature vectors $\hat{\mathbf{Y}}_{\mathsf{mid}}^{(i)}\in\mathbb{R}^{T\times D_{k^{(\mathsf{a})}}}$ conditioned on $\eta^{(\mathsf{a})}$, $\hat{\eta}_{\mathsf{Rx}}^{(i)}$ and $k^{(i)}$:
\begin{equation}
    \hat{\mathbf{Y}}_{\mathsf{mid}}^{(i)}
    =
    g_{\boldsymbol{\phi}_{\mathsf{inner}}}
    \big(
        \hat{\mathbf{Y}}^{(i)},
        \eta^{(\mathsf{a})},
        \hat{\eta}_{\mathsf{Rx}}^{(i)}, k^{(i)}
    \big).
\end{equation}

After processing all $U$ transmission blocks sequentially with the inner decoder, the intermediate feature vectors $\{\hat{\mathbf{Y}}_{\mathsf{mid}}^{(i)}\}_{i=1}^{U}$ are concatenated to form the intermediate feature vector sequence $\hat{\mathbf{Y}}_{\mathsf{mid}}=\big[(\hat{\mathbf{Y}}_{\mathsf{mid}}^{(1)})^{\top},\dots,(\hat{\mathbf{Y}}_{\mathsf{mid}}^{(U)})^{\top}\big]^{\top}\in\mathbb{R}^{N\times D_{k^{(\mathsf{a})}}}$.

Finally, the outer decoder $g_{\boldsymbol{\phi}_{\mathsf{outer}}}$ reconstructs the source image $\hat{\mathbf{X}}$ from $\hat{\mathbf{Y}}_{\mathsf{mid}}$ conditioned on $\eta^{(\mathsf{a})}$:
\begin{equation}
\label{eq:outer_decoding}
    \hat{\mathbf{X}}
    =
    g_{\boldsymbol{\phi}_{\mathsf{outer}}}
    \big(
        \hat{\mathbf{Y}}_{\mathsf{mid}},
        \eta^{(\mathsf{a})}
    \big).
\end{equation}

\begin{algorithm}[!t]
\small
\caption{Communication Process of euJSCC}
\label{alg:euJSCC-Forward-FAST}
\begin{algorithmic}[1]
\State \textbf{Input: }Source image $\mathbf{X}$, average SNR $\eta^{(\mathsf{a})}$, block-wise CQIs $\{\hat{\eta}_{\mathsf{Tx}}^{(i)}\}_{i=1}^{U}$ at the transmitter and $\{\hat{\eta}_{\mathsf{Rx}}^{(i)}\}_{i=1}^{U}$ at the receiver, modulation orders $\{m_{k^{(i)}}\}_{i=1}^{U}$ and corresponding indices $\{k^{(i)}\}_{i=1}^{U}$, data symbol length per transmission block $T$, outer encoder--decoder $(f_{\boldsymbol{\theta}_{\mathsf{outer}}},g_{\boldsymbol{\phi}_{\mathsf{outer}}})$, inner encoder--decoder $(f_{\boldsymbol{\theta}_{\mathsf{inner}}},g_{\boldsymbol{\phi}_{\mathsf{inner}}})$, DCG network $\mathcal{C}_{\boldsymbol{\Psi}}$
\State Encode $\mathbf{X}$ into $\mathbf{Y}_{\mathsf{mid}}$ with $\eta^{(\mathsf{a})}$ via $f_{\boldsymbol{\theta}_{\mathsf{outer}}}$
\State Partition $\mathbf{Y}_{\mathsf{mid}}$ into $\{\mathbf{Y}_{\mathsf{mid}}^{(i)}\}_{i=1}^{U}$
\For{$i\gets1$ to $U$}
    \State Adapt $\mathbf{Y}_{\mathsf{mid}}^{(i)}$ into $\mathbf{Y}^{(i)}$ with $\eta^{(\mathsf{a})}$, $\hat{\eta}_{\mathsf{Tx}}^{(i)}$ and $k^{(i)}$ via $f_{\boldsymbol{\theta}_{\mathsf{inner}}}$
    \State Generate $\mathbf{C}_{\mathsf{Tx}}^{(i)}$ with $\hat{\eta}_{\mathsf{Tx}}^{(i)}$ and $k^{(i)}$ via $\mathcal{C}_{\boldsymbol{\Psi}}$
    \State Quantize $\mathbf{Y}^{(i)}$ into $\mathbf{z}^{(i)}$ using $\mathbf{C}_{\mathsf{Tx}}^{(i)}$
    \State Modulate $\mathbf{z}^{(i)}$ into $\mathbf{s}^{(i)}$ via $m_{k^{(i)}}$-ary modulation
    \State Transmit $\mathbf{s}^{(i)}$ over the channel and obtain $\hat{\mathbf{s}}^{(i)}$
    \State Equalize $\hat{\mathbf{s}}^{(i)}$ into $\tilde{\mathbf{s}}^{(i)}$ via ZF equalization
    \State Demodulate $\tilde{\mathbf{s}}^{(i)}$ into $\hat{\mathbf{z}}^{(i)}$ via $m_{k^{(i)}}$-ary demodulation
    \State Generate $\mathbf{C}_{\mathsf{Rx}}^{(i)}$ with $\hat{\eta}_{\mathsf{Rx}}^{(i)}$ and $k^{(i)}$ via $\mathcal{C}_{\boldsymbol{\Psi}}$
    \State Dequantize $\hat{\mathbf{z}}^{(i)}$ into $\hat{\mathbf{Y}}^{(i)}$ using $\mathbf{C}_{\mathsf{Rx}}^{(i)}$
    \State Adapt $\hat{\mathbf{Y}}^{(i)}$ into $\hat{\mathbf{Y}}_{\mathsf{mid}}^{(i)}$ with $\eta^{(\mathsf{a})}$, $\hat{\eta}_{\mathsf{Rx}}^{(i)}$ and $k^{(i)}$ via $g_{\boldsymbol{\phi}_{\mathsf{inner}}}$
\EndFor
\State Aggregate $\{\hat{\mathbf{Y}}_{\mathsf{mid}}^{(i)}\}_{i=1}^{U}$ to reconstruct $\hat{\mathbf{Y}}_{\mathsf{mid}}$
\State Reconstruct $\hat{\mathbf{X}}$ from $\hat{\mathbf{Y}}_{\mathsf{mid}}$ with $\eta^{(\mathsf{a})}$ via $g_{\boldsymbol{\phi}_{\mathsf{outer}}}$
\State \textbf{Output: }$\hat{\mathbf{X}}$
\end{algorithmic}
\end{algorithm}

\subsection{End-to-End Training Objective}
\label{subsec:End-to-End Training Objective}

To enable end-to-end training despite the non-differentiable quantization step, we adopt the VQ-VAE framework~\cite{van2017neural} with a straight-through estimator (STE). Formally, the overall training objective is defined as:
\begin{align}
\label{eq:VQVAE LOSS}
\begin{split}
    \mathcal{L}
    =
    \mathbb{E}_{\mathbf{X}}
    [\mathsf{MSE}(\hat{\mathbf{X}},\mathbf{X})
    &+\frac{1}{U}
    \sum_{i=1}^{U}
    \alpha_{k^{(i)}}\cdot
    \mathsf{MSE}
    (\hat{\mathbf{Y}}^{(i)},\mathbf{Y}_{\mathsf{d}}^{(i)})
    \\
    &+\frac{1}{U}
    \sum_{i=1}^{U}
    \beta_{k^{(i)}}\cdot
    \mathsf{MSE}
    (\mathbf{Y}^{(i)},\hat{\mathbf{Y}}_{\mathsf{d}}^{(i)})],
\end{split}
\end{align}
where $\mathsf{MSE}$ denotes the mean squared error function, and $\mathbf{Y}_{\mathsf{d}}^{(i)}$ and $\hat{\mathbf{Y}}_{\mathsf{d}}^{(i)}$ are detached copies of $\mathbf{Y}^{(i)}$ and $\hat{\mathbf{Y}}^{(i)}$, respectively, used to block gradient propagation. Given a fixed set of coefficient pairs $\{(\alpha_k,\beta_k)\}_{k=1}^{K}$, the block-wise coefficients $\{(\alpha_{k^{(i)}},\beta_{k^{(i)}})\}_{i=1}^{U}$ act as hyperparameters that control the relative weight of each loss component corresponding to $k^{(i)}$. The first term represents the image reconstruction loss, ensuring pixel-level fidelity. The second term, the codebook loss, aligns the codebook embeddings with the encoder feature vectors assigned to them, while the third term, the commitment loss, constrains the encoder outputs to remain close to their corresponding codewords. Together, these two losses mitigate the mismatch between $\hat{\mathbf{Y}}^{(i)}$ and $\mathbf{Y}^{(i)}$, thereby enabling stable and consistent end-to-end training with the STE despite the discrete nature of the quantization step. Since the scales of the codebook and commitment losses vary with the codebook dimension $D_k$ and codebook size $m_k$, different $(\alpha_k,\beta_k)$ values are applied for each modulation order.

\vspace{-2mm}
\section{Extended Universal Joint Source-Channel Coding}
\label{sec:Extended Universal Joint Source-Channel Coding}

The proposed euJSCC framework adapts feature encoding/decoding and codebook generation to channel conditions through two key components: (i) an inner--outer encoder--decoder for parameter-efficient processing of feature vectors with dynamically varying channel dimensions, and (ii) a DCG network for generating corresponding codebooks. Both components adapt to the available CQI and selected modulation order. We next describe these components and the hypernetwork-based normalization layer that supports their adaptation to modulation orders and fine-grained channel conditions.

\vspace{-3mm}
\subsection{Implementation of the Encoder-Decoder Architecture}
\label{subsec:Implementation of the Encoder-Decoder Architecture}

\begin{figure}
    \centering
    \includegraphics[width=0.8\linewidth]{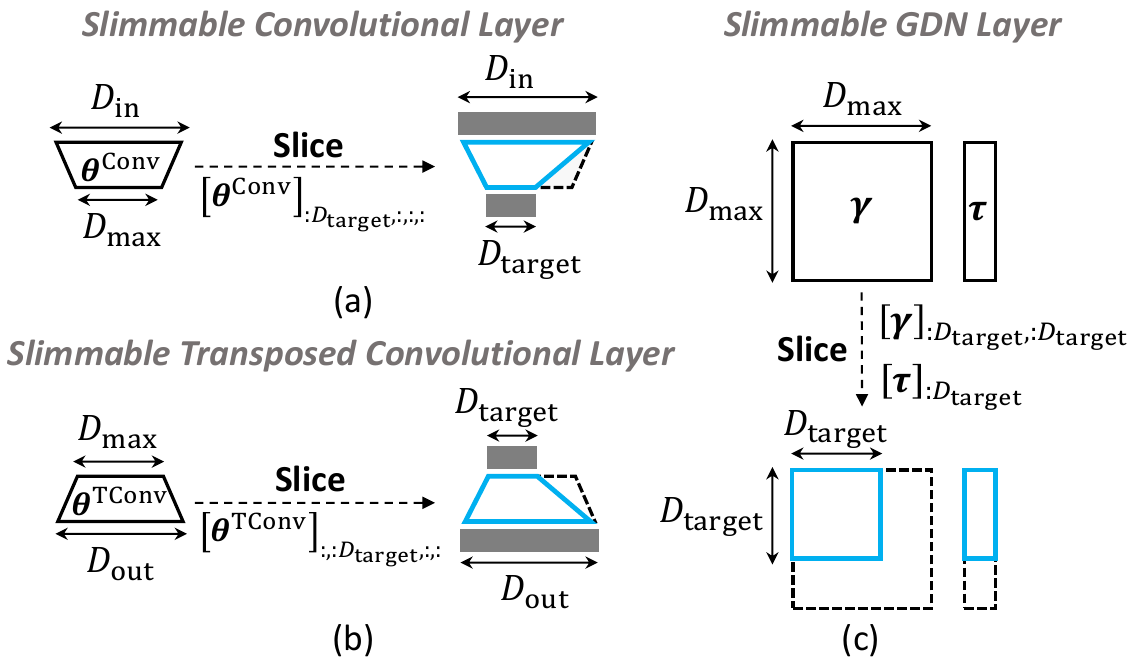}
    \vspace{-3mm}
    \caption{An illustration of slimmable operation on (a) convolutional layer, (b) transposed convolutional layer, and (c) GDN layer.}
    \label{fig:slimmable}
    \vspace{-5mm}
\end{figure}

\begin{figure*}[ht!]
    \centering
    \includegraphics[width=0.8\linewidth]{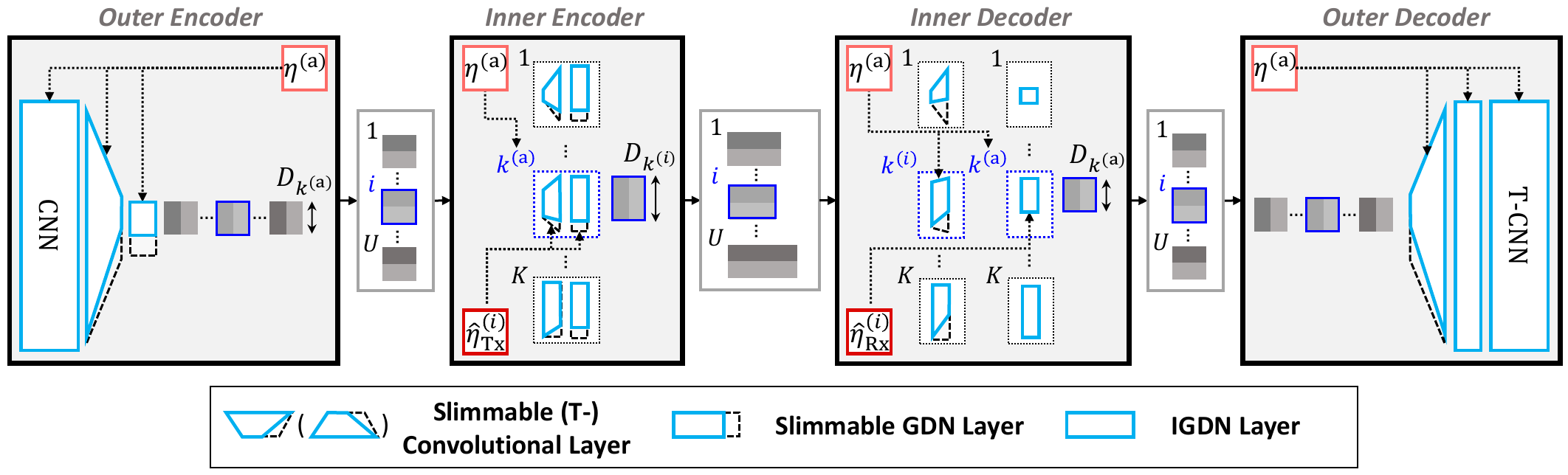}
    \vspace{-3mm}
    \caption{The overall euJSCC encoder--decoder architecture, decomposed into inner and outer modules. ``CNN'', ``T-CNN'', and ``Slimmable T-Convolutional Layer'' denote CNN blocks, transposed CNN blocks, and slimmable transposed convolutional layer, respectively. The blue-highlighted path shows an example processing flow for the $i$-th transmission block.}
    \label{fig:codec}
    \vspace{-5mm}
\end{figure*}

In practical wireless systems, the channel condition can vary over short time scales. To address this situation, \cite{li2025coarse} proposed a two-phase encoding-decoding process consisting of a coarse-grained phase, which uses the average SNR computed from long-term channel statistics, and a fine-grained phase, which handles rapid short-term channel fluctuations. Motivated by this idea, we integrate a similar strategy into the design of our VQ-JSCC encoder--decoder architecture for digital semantic communication, enabling robust and efficient operation under time-varying channel conditions.

The JSCC encoder $\mathcal{F}_{\boldsymbol{\Theta}}$ and decoder $\mathcal{G}_{\boldsymbol{\Phi}}$, introduced in Sections~\ref{subsubsec:Adaptive Semantic Encoding and Vector Quantization} and~\ref{subsubsec:Adaptive Vector Dequantization and Semantic Decoding}, are implemented as stacks of (transposed) convolutional neural network (CNN) blocks. Each block consists of a (transposed) convolutional layer followed by a (inverse) generalized divisive normalization (GDN) layer, which are widely adopted in JSCC models for their ability to capture inter-channel dependencies of the output tensor channels from each neural network (NN) layer. Additionally, both the inner encoder $f_{\boldsymbol{\theta}_{\mathsf{inner}}}$ and inner decoder $g_{\boldsymbol{\phi}_{\mathsf{inner}}}$ are implemented with parallel instances of a single convolutional layer and a (inverse) GDN layer to enable low-complexity, block-wise refinement. The standard (inverse) GDN operations are introduced in \textbf{Remark~\ref{remark:gdn_igdn}}.

\begin{remark}[Standard GDN and inverse GDN]
\label{remark:gdn_igdn}
Let $\mathbf{t}_{n}$ denote the output tensor with $D_{\mathsf{in}}$ tensor channel dimensions from the $n$-th layer, and $t_{n,d,p}$ denote the $p$-th element in the $d$-th tensor channel of $\mathbf{t}_{n}$. The standard GDN operation produces the normalized tensor $\tilde{\mathbf{t}}_{n}$ as:
\begin{equation}
\label{eq:standardGDN}
    \tilde{t}_{n,d,p}
    =
    \frac{t_{n,d,p}}
    {\sqrt{\tau_d + \sum_{d'=1}^{D_{\mathsf{in}}} \gamma_{d',d}\, t^2_{n,d',p}}},
    \quad d\in[1{:}D_{\mathsf{in}}],
\end{equation}
where $\boldsymbol{\tau} \in \mathbb{R}^{D_{\mathsf{in}}}$ is a bias vector and $\boldsymbol{\gamma} \in \mathbb{R}^{D_{\mathsf{in}}\times D_{\mathsf{in}}}$ is a coefficient matrix. Symmetrically, the inverse GDN (IGDN) operation performs the inverse operation using its own set of parameters:
\begin{equation}
\label{eq:standardIGDN}
    \tilde{t}_{n,d,p}
    =
    t_{n,d,p}
    \sqrt{\tau_{d} + \sum_{d'=1}^{D_{\mathsf{in}}} \gamma_{d',d}\, t^{2}_{n,d',p}},
    \quad d\in[1{:}D_{\mathsf{in}}].
\end{equation}
\end{remark}

Before detailing the specific architectures for $\mathcal{F}_{\boldsymbol{\Theta}}$ and $\mathcal{G}_{\boldsymbol{\Phi}}$, it is important to introduce the \textit{slimmable operations} used to handle the varying feature vector channel dimensions $D_k \in \{D_1, \dots,D_K\}$ that arise from adapting to different modulation orders. Standard NN layers require fixed input and output tensor channel dimensions. However, in a VQ-JSCC system that should support multiple dimensions according to the current modulation order condition, maintaining separate networks for each possible input or output channel dimension significantly increases system and memory complexity. A slimmable operation~\cite{yu2018slimmable} enables each NN layer type to reuse a single set of weights while processing or producing tensors with different channel dimensions. This mechanism is one of the key components of our parameter-efficient design, and its application to three types of NN layer utilized in $\mathcal{F}_{\boldsymbol{\Theta}}$ and $\mathcal{G}_{\boldsymbol{\Phi}}$ is described below.

\begin{itemize}

\item \textbf{Slimmable Convolutional Layer~\cite{yu2018slimmable}:}
Let $\boldsymbol{\theta}^{\mathsf{Conv}} \in \mathbb{R}^{D_{\mathsf{max}} \times D_{\mathsf{in}} \times \kappa \times \kappa}$ denote the weight tensor of a 2D convolutional layer with kernel size $\kappa \times \kappa$ (or $\boldsymbol{\theta}^{\mathsf{Conv}} \in \mathbb{R}^{D_{\mathsf{max}} \times D_{\mathsf{in}} \times \kappa}$ for the 1D case). Here, $D_{\mathsf{in}}$ is the input tensor channel dimension and $D_{\mathsf{max}}$ is the maximum output tensor channel dimension supported by the layer. In a standard convolutional layer, all $D_{\mathsf{max}}$ filters, the shape of which is $D_{\mathsf{in}} \times \kappa \times \kappa$ (or $D_{\mathsf{in}} \times \kappa$ for the 1D case), are applied to the input tensor to produce an output tensor with channel dimension $D_{\mathsf{max}}$. In order to generate an output tensor with varying channel dimension $D_{\mathsf{target}}$, as shown in Fig.~\ref{fig:slimmable}(a), the same weight tensor is shared, but only the first $D_{\mathsf{target}}$ filters, i.e., $[\boldsymbol{\theta}^{\mathsf{Conv}}]_{:D_{\mathsf{target}},:,:,:}$ (or $[\boldsymbol{\theta}^{\mathsf{Conv}}]_{:D_{\mathsf{target}},:,:}$ for the 1D case), are used.

\item \textbf{Slimmable Transposed Convolutional Layer~\cite{yu2018slimmable}:}
Let $\boldsymbol{\theta}^{\mathsf{TConv}} \in \mathbb{R}^{D_{\mathsf{out}} \times D_{\mathsf{max}} \times \kappa \times \kappa}$ denote the weight tensor of a 2D transposed convolutional layer with kernel size $\kappa \times \kappa$ (or $\boldsymbol{\theta}^{\mathsf{TConv}} \in \mathbb{R}^{D_{\mathsf{out}} \times D_{\mathsf{max}} \times \kappa}$ for the 1D case). Here, $D_{\mathsf{max}}$ is the maximum input tensor channel dimension supported by the layer and $D_{\mathsf{out}}$ is the output tensor channel dimension. In contrast to the convolutional layer, as shown in Fig.~\ref{fig:slimmable}(b), when the input tensor has varying channel dimension $D_{\mathsf{target}}$, the sliced weight tensor $[\boldsymbol{\theta}^{\mathsf{TConv}}]_{:,:D_{\mathsf{target}},:,:}$ (or $[\boldsymbol{\theta}^{\mathsf{TConv}}]_{:,:D_{\mathsf{target}},:}$ for the 1D case) is applied to the input tensor to produce an output tensor with channel dimension $D_{\mathsf{out}}$.

\item \textbf{Slimmable GDN Layer:}
Based on the standard GDN operation defined in \eqref{eq:standardGDN}, we apply the slimmable operations to its normalization parameters. The full parameters for input tensors with channel dimension $D_{\mathsf{max}}$, i.e., $\boldsymbol{\tau} \in \mathbb{R}^{D_{\mathsf{max}}}$ and $\boldsymbol{\gamma} \in \mathbb{R}^{D_{\mathsf{max}}\times D_{\mathsf{max}}}$ are stored. Fig.~\ref{fig:slimmable}(c) shows that when the input tensor has a channel dimension $D_{\mathsf{target}} \le D_{\mathsf{max}}$, the layer uses only the leading sub-vector $[\boldsymbol{\tau}]_{:D_{\mathsf{target}}}$ and the leading sub-matrix $[\boldsymbol{\gamma}]_{:D_{\mathsf{target}},:D_{\mathsf{target}}}$ to perform the GDN operation.

\end{itemize}

Now, we describe how these slimmable operations are integrated into the specific encoder-decoder architecture. The overall process is illustrated in Fig.~\ref{fig:codec}.

\textbf{Outer Encoder ${f}_{\boldsymbol{\theta}_{\mathsf{outer}}}$:}
${f}_{\boldsymbol{\theta}_{\mathsf{outer}}}$ generates an intermediate feature vector sequence $\mathbf{Y}_{\mathsf{mid}}$ with channel dimension $D_{k^{(\mathsf{a})}} \in \{D_1,\dots,D_K\}$ determined by $k^{(\mathsf{a})}$. To support this variable-dimensional output, the final CNN block of ${f}_{\boldsymbol{\theta}_{\mathsf{outer}}}$ uses a single slimmable convolutional layer followed by a slimmable GDN layer.

\textbf{Inner Encoder ${f}_{\boldsymbol{\theta}_{\mathsf{inner}}}$:}
The partitioned feature vectors $\mathbf{Y}_{\mathsf{mid}}^{(i)}$ are processed by ${f}_{\boldsymbol{\theta}_{\mathsf{inner}}}$. ${f}_{\boldsymbol{\theta}_{\mathsf{inner}}}$ receives feature vectors with channel dimension $D_{k^{(\mathsf{a})}}$ and produces output feature vectors with channel dimension $D_{k^{(i)}}\in\{D_1,\dots,D_K\}$ determined by $k^{(i)}$. Therefore, the module is implemented using $K$ parallel 1D slimmable convolutional layers followed by slimmable GDN layers. Conditioned on $\eta^{(\mathsf{a})}$, $\eta^{(i)}_{\mathsf{Tx}}$ and $k^{(i)}$, the $k^{(\mathsf{a})}$-th layer is activated to process the input $\mathbf{Y}^{(i)}_{\mathsf{mid}}$ with input channel dimension $D_{k^{(\mathsf{a})}}$, and ${f}_{\boldsymbol{\theta}_{\mathsf{inner}}}$ adaptively produces $\mathbf{Y}^{(i)}$ with the target dimension $D_{k^{(i)}}$ specified by $k^{(i)}$.

\textbf{Inner Decoder $g_{\boldsymbol{\phi}_{\mathsf{inner}}}$:}
Symmetrically, $g_{\boldsymbol{\phi}_{\mathsf{inner}}}$ adapts the received and dequantized feature vectors $\hat{\mathbf{Y}}^{(i)}$ with channel dimension $D_{k^{(i)}}$ into feature vectors with channel dimension $D_{k^{(\mathsf{a})}}\in\{D_1,\dots,D_K\}$. Specifically, the module is implemented using $K$ parallel 1D slimmable convolutional layers followed by IGDN layers. Conditioned on $\eta^{(\mathsf{a})}$, $\eta^{(i)}_{\mathsf{Rx}}$ and $k^{(i)}$, the $k^{(i)}$-th slimmable convolutional layer and the $k^{(\mathsf{a})}$-th IGDN layer are activated to process the input $\hat{\mathbf{Y}}^{(i)}$ with channel dimension $D_{k^{(i)}}$ and adaptively produce the output with channel dimension $D_{k^{(\mathsf{a})}}$.

\textbf{Outer Decoder ${g}_{\boldsymbol{\phi}_{\mathsf{outer}}}$:}
After concatenating all refined block-wise feature vectors into $\hat{\mathbf{Y}}_{\mathsf{mid}}$, ${g}_{\boldsymbol{\phi}_{\mathsf{outer}}}$ decodes $\hat{\mathbf{Y}}_{\mathsf{mid}}$ to reconstruct $\hat{\mathbf{X}}$. To accommodate the variable channel dimension $D_{k^{(\mathsf{a})}}$, the first transposed CNN block uses a slimmable transposed convolutional layer followed by an IGDN layer.

\begin{remark}[Case of $N\neq TU$]
In this case, the definitions for the final $U$-th block are modified accordingly. While the first $U-1$ blocks are processed as described above, the inner operations for the final block are given by:
\begin{align}
    \mathbf{Y}_{\mathsf{mid}}^{(U)}
    &=
    [\mathbf{y}_{\mathsf{mid},(U-1)T+1},\dots,
    \mathbf{y}_{\mathsf{mid},N}]^{\top}
    \in
    \mathbb{R}^{T_U\times D_{k^{(\mathsf{a})}}},
    \\
    \mathbf{Y}^{(U)}
    &=
    {f}_{\boldsymbol{\theta}_{\mathsf{inner}}}
    \big(
        \mathbf{Y}^{(U)}_{\mathsf{mid}},\eta^{(\mathsf{a})},
        \hat{\eta}_{\mathsf{Tx}}^{(U)}, k^{(U)}
    \big)
    \in
    \mathbb{R}^{T_U\times D_{k^{(U)}}},
    \\
    \hat{\mathbf{Y}}^{(U)}_{\mathsf{mid}}
    &=
    {g}_{\boldsymbol{\phi}_{\mathsf{inner}}}
    \big(
        \hat{\mathbf{Y}}^{(U)},\eta^{(\mathsf{a})},
        \hat{\eta}_{\mathsf{Rx}}^{(U)}, k^{(U)}
    \big)
    \in
    \mathbb{R}^{T_U\times D_{k^{(\mathsf{a})}}},
\end{align}
where $T_U=N-(U-1)T$ is the length of the final block and $\hat{\mathbf{Y}}^{(U)}\in\mathbb{R}^{T_U\times D_{k^{(U)}}}$ is the corresponding dequantized feature vectors at the receiver. Concatenating $\{\hat{\mathbf{Y}}^{(i)}_{\mathsf{mid}}\}_{i=1}^{U}$ then correctly reconstructs $\hat{\mathbf{Y}}_{\mathsf{mid}}\in\mathbb{R}^{N\times D_{k^{(\mathsf{a})}}}$.
\end{remark}

\vspace{-3mm}
\subsection{Dynamic Codebook Generator}
\label{subsec:Dynamic Codebook Generator}

\begin{figure*}[ht!]
    \centering
    \includegraphics[width=0.8\linewidth]{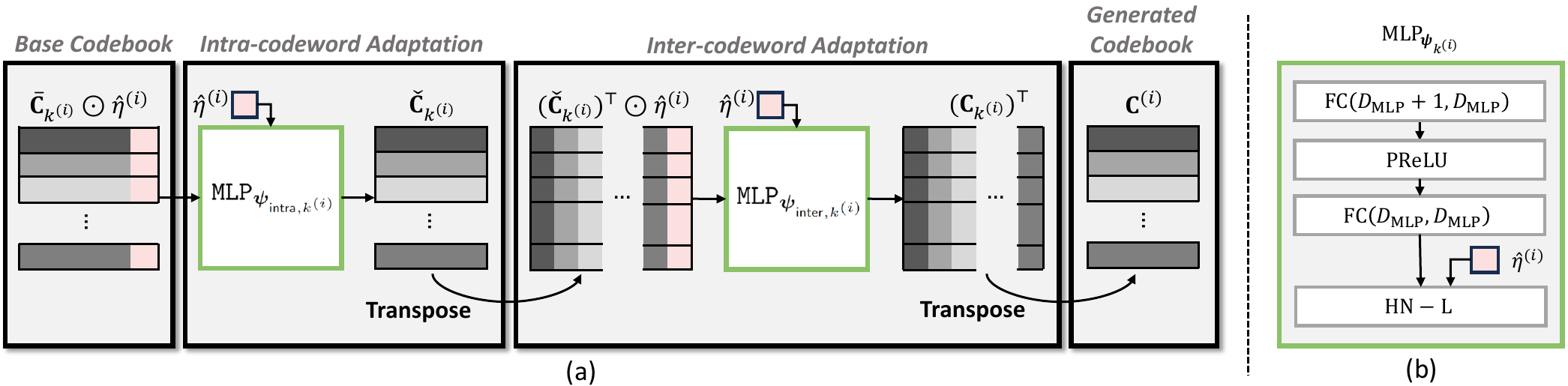}
    \vspace{-3mm}
    \caption{An illustration of the proposed DCG network, which dynamically generates a codebook conditioned on $\hat{\eta}^{(i)}$ and $k^{(i)}$. For each modulation order $k$, a base codebook $\bar{\mathbf{C}}_{k}$, an intra-codeword adaptation $\mathtt{MLP}_{\boldsymbol{\psi}_{\mathsf{intra},k^{(i)}}}$, and an inter-codeword adaptation $\mathtt{MLP}_{\boldsymbol{\psi}_{\mathsf{inter},k^{(i)}}}$ are separately maintained, and the components corresponding to the given $k=k^{(i)}$ are activated. Both $\mathtt{MLP}_{\boldsymbol{\psi}_{\mathsf{intra},k^{(i)}}}$ and $\mathtt{MLP}_{\boldsymbol{\psi}_{\mathsf{inter},k^{(i)}}}$ follow the structure $\mathtt{MLP}_{\boldsymbol{\psi}_{k^{(i)}}}$ shown in (b), where $D_{\mathtt{MLP}}=D_{k^{(i)}}$ for the intra-codeword adaptation MLP and $D_{\mathtt{MLP}}=m_{k^{(i)}}$ for the inter-codeword adaptation MLP.}
    \label{fig:DCG}
    \vspace{-5mm}
\end{figure*}

Since the operations in Sections~\ref{subsec:Dynamic Codebook Generator} and~\ref{subsec:Hypernetwork-based Normalization Layer} are performed at both the transmitter and receiver, we omit the transmitter/receiver subscripts for notational simplicity. For example, $\hat{\eta}^{(i)}$ denotes the CQI available at the corresponding side. In VQ-JSCC, adapting feature vectors to the wireless channel condition alone is insufficient. A proper alignment of the VQ codebook to the extracted feature vectors is also crucial. Moreover, since the optimal codebook varies with fine-grained channel conditions, using a fixed codebook for each modulation order can lead to significant mismatches between the feature vectors produced by the JSCC encoder at the transmitter and the codewords detected at the receiver. Such mismatches degrade performance by increasing feature distortion errors.

To address these challenges, we propose the DCG network, which adaptively transforms a base codebook for each modulation order into a CQI-specific codebook. As illustrated in Fig.~\ref{fig:DCG}(a), the DCG network comprises three key components that operate according to $\hat{\eta}^{(i)}$ and $k^{(i)}$.
\begin{enumerate}
    \item \textbf{Base Codebook:}
    For each modulation order $k^{(i)}\in[1{:}K]$, we employ a base codebook $\bar{\mathbf{C}}_{k^{(i)}} = [\bar{\mathbf{c}}_{k^{(i)},1}, \dots, \bar{\mathbf{c}}_{k^{(i)},m_{k^{(i)}}}]^\top\in\mathbb{R}^{m_{k^{(i)}}\times D_{k^{(i)}}},$ where $\bar{\mathbf{c}}_{k^{(i)},j}\in\mathbb{R}^{D_{k^{(i)}}}$ is the $j$-th base codeword for the $k^{(i)}$-th modulation order.

    \item \textbf{Intra-codeword Adaptation:}
    Since the base codebooks are not optimal for $\hat{\eta}^{(i)}$, we refine each base codeword with a CQI-conditioned transformation. Specifically, each base codeword $\bar{\mathbf{c}}_{k^{(i)},j}\in\mathbb{R}^{D_{k^{(i)}}}$ is concatenated with $\hat{\eta}^{(i)}$ and passed through an intra-codeword adaptation MLP $\mathtt{MLP}_{\boldsymbol{\psi}_{\mathsf{intra},k^{(i)}}}$, parameterized by $\boldsymbol{\psi}_{\mathsf{intra},k^{(i)}}$, as
    \begin{align}
        \check{\mathbf{c}}_{k^{(i)},j}
        =
        \mathtt{MLP}_{\boldsymbol{\psi}_{\mathsf{intra},k^{(i)}}}
        \big(
            \bar{\mathbf{c}}_{k^{(i)},j}
            \odot
            \hat{\eta}^{(i)}
        \big),
    \end{align}
    where $\odot$ denotes the concatenation operation along the column dimensions. $\mathtt{MLP}_{\boldsymbol{\psi}_{\mathsf{intra}, k^{(i)}}}$ is composed of two fully connected layers followed by a layer normalization (LN) layer, with hidden dimension $D_{\mathtt{MLP}}=D_{k^{(i)}}$ in Fig.~\ref{fig:DCG}(b). Applying this adaptation to all $m_{k^{(i)}}$ codewords yields the intra-adapted codebook $\check{\mathbf{C}}_{k^{(i)}} = [\check{\mathbf{c}}_{k^{(i)},1}, \dots, \check{\mathbf{c}}_{k^{(i)},m_{k^{(i)}}}]^{\top}\in\mathbb{R}^{m_{k^{(i)}}\times D_{k^{(i)}}}$.
    
    \item \textbf{Inter-codeword Adaptation:}
    While intra-codeword adaptation adjusts each codeword individually, it does not consider the relative geometry among codewords. This geometric structure is particularly critical at low SNRs for each modulation, where detection errors frequently occur. If neighboring constellation symbols are mapped to codewords that are far apart in feature space, small detection errors on the constellation map can cause large performance degradation at the decoder. To address this, we apply inter-codeword adaptation, refining the codebook column-wise after intra-codeword adaptation. Let $[\check{\mathbf{C}}_{k^{(i)}}]_{:,d}\in\mathbb{R}^{m_{k^{(i)}}}$ denote the $d$-th column of the intra-adapted codebook, which collects the $d$-th component across all $m_{k^{(i)}}$ codewords. Similarly, $[\check{\mathbf{C}}_{k^{(i)}}]_{:,d}$ is concatenated with $\hat{\eta}^{(i)}$ and then processed by the inter-codeword adaptation MLP $\mathtt{MLP}_{\boldsymbol{\psi}_{\mathsf{inter},k^{(i)}}}$:
    \begin{align}
        [\mathbf{C}_{k^{(i)}}]_{:,d}
        =
        \mathtt{MLP}_{\boldsymbol{\psi}_{\mathsf{inter},k^{(i)}}}
        \big(
            [\check{\mathbf{C}}_{k^{(i)}}]_{:,d}
            \odot
            \hat{\eta}^{(i)}
        \big),
    \end{align}
    where the MLP architecture is identical to that of $\mathtt{MLP}_{\boldsymbol{\psi}_{\mathsf{intra},k^{(i)}}}$, except that $D_{\mathtt{MLP}}=m_{k^{(i)}}$. Stacking the outputs across all columns produces the final codebook for $\hat{\eta}^{(i)}$ and $m_{k^{(i)}}$, i.e., $\mathbf{C}^{(i)} \in \mathbb{R}^{m_{k^{(i)}}\times D_{k^{(i)}}}.$
\end{enumerate}

The overall DCG network can be viewed as a unified network $\mathcal{C}_{\boldsymbol{\Psi}}$, parameterized by $\boldsymbol{\Psi}=\{\boldsymbol{\psi}_{\mathsf{intra},k}, \boldsymbol{\psi}_{\mathsf{inter},k}, \bar{\mathbf{C}}_k\}_{k=1}^K$, where, for the $i$-th transmission block with modulation order $k^{(i)}$, only the associated parameters, i.e., $\{\boldsymbol{\psi}_{\mathsf{intra},k^{(i)}}, \boldsymbol{\psi}_{\mathsf{inter},k^{(i)}}, \bar{\mathbf{C}}_{k^{(i)}}\}$, are activated to generate an adaptive codebook $\mathbf{C}^{(i)}=\mathcal{C}_{\boldsymbol{\Psi}}(\hat{\eta}^{(i)},k^{(i)})$. In fact, this design is analogous to the design principle of MLP-Mixer architectures~\cite{tolstikhin2021mlp}, where feature vector processing is decomposed into intra-feature and inter-feature transformations. It is realized by lightweight MLPs, which effectively achieve global expressivity comparable to much more computationally complex attention mechanisms of transformers. In a similar manner, our DCG network leverages base codebooks in conjunction with compact MLP layers to produce codebooks specialized for $\hat{\eta}^{(i)}$ and $k^{(i)}$, achieving fine-grained adaptation with high efficiency in both parameterization and computation.

\vspace{-3mm}
\subsection{Hypernetwork-based Normalization Layer}
\label{subsec:Hypernetwork-based Normalization Layer}

\begin{figure}
    \centering
    \includegraphics[width=0.8\linewidth]{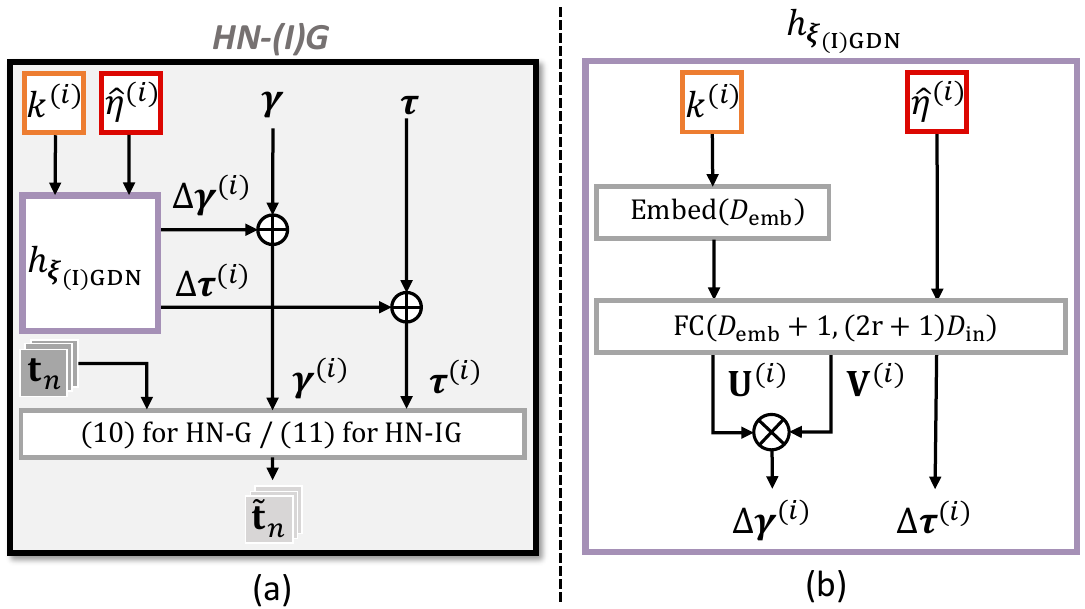}
    \vspace{-3mm}
    \caption{(a) Adaptation process for (I)GDN parameters of the $\mathtt{HN\text{--}(I)G}$ layer. (b) Network architecture of hypernetwork $h_{\boldsymbol{\xi}_{\mathsf{(I)GDN}}}$ in the $\mathtt{HN\text{--}(I)G}$ layer.}
    \label{fig:HN}
    \vspace{-5mm}
\end{figure}

Normalization layers significantly facilitate the training of NN models and enhance their performance by stabilizing the statistics of an output tensor from each NN layer. As established in uJSCC~\cite{huh2025universal}, the adaptability of VQ-JSCC networks that support different modulation orders and codebooks is significantly improved through separate switchable normalization layers for each modulation order. However, since it is required to adapt codebooks according to the block-wise CQI within a given modulation order, this approach is ill-suited for our framework, leading to prohibitive parameter overhead if naively extended.

To overcome this limitation, we adopt a hypernetwork\footnote{A hypernetwork is a small auxiliary network that produces the parameters of another network based on conditioning variables.}-based approach inspired by adaptive instance normalization (AdaIN)~\cite{huang2017arbitrary}, where normalization parameters are dynamically generated from conditioning variables. Specifically, instead of storing excessive sets of normalization parameters, we employ a lightweight hypernetwork that produces normalization parameters conditioned on the channel condition supplied to each module and the modulation order index. This principle is applied to adapt the (I)GDN layers in the JSCC encoder--decoder in Section~\ref{subsec:Implementation of the Encoder-Decoder Architecture} and the LN layers of the DCG network in Section~\ref{subsec:Dynamic Codebook Generator}, as detailed below.

\subsubsection{Hypernetwork-based (I)GDN Layer}

The standard GDN operation in~\eqref{eq:standardGDN} relies on fixed trainable parameters: a bias vector $\boldsymbol{\tau}$ and a coefficient matrix $\boldsymbol{\gamma}$. We first describe the $\mathtt{HN\text{-}(I)G}$ operation for the $i$-th transmission block. To condition the block-wise GDN operation on $\hat{\eta}^{(i)}$ and $k^{(i)}$, we introduce a hypernetwork that generates additive offsets $\Delta\boldsymbol{\tau}^{(i)} \in \mathbb{R}^{D_{\mathsf{in}}}$ and $\Delta\boldsymbol{\gamma}^{(i)} \in \mathbb{R}^{D_{\mathsf{in}}\times D_{\mathsf{in}}}$ based on $\hat{\eta}^{(i)}$ and $k^{(i)}$. As illustrated in Fig.~\ref{fig:HN}(a), the adapted parameters for the $i$-th transmission block are $\boldsymbol{\tau}^{(i)} = \boldsymbol{\tau} + \Delta\boldsymbol{\tau}^{(i)}$ and $\boldsymbol{\gamma}^{(i)} = \boldsymbol{\gamma} + \Delta\boldsymbol{\gamma}^{(i)}$. The offsets are generated by a hypernetwork $h_{\boldsymbol{\xi}_{\mathsf{GDN}}}$, which embeds $k^{(i)}$, concatenates it with $\hat{\eta}^{(i)}$, and maps the result through a single fully connected layer.

However, directly generating $\Delta\boldsymbol{\gamma}^{(i)}$ requires an output size of $\mathcal{O}(D_{\mathsf{in}}^2)$, resulting in an excessive number of parameters in $h_{\boldsymbol{\xi}_{\mathsf{GDN}}}$, which can be comparable to the entire JSCC model. To mitigate this overhead, we adopt the low-rank adaptation (LoRA) method~\cite{hu2022lora} and decompose $\Delta\boldsymbol{\gamma}^{(i)} = \mathbf{U}^{(i)}\mathbf{V}^{(i)}$, where $\mathbf{U}^{(i)} \in \mathbb{R}^{D_{\mathsf{in}}\times r}$ and $\mathbf{V}^{(i)} \in \mathbb{R}^{r\times D_{\mathsf{in}}}$ with $r \ll D_{\mathsf{in}}$. Accordingly, as shown in Fig.~\ref{fig:HN}(b), the hypernetwork outputs
\begin{equation}
    \mathbf{U}^{(i)}, \mathbf{V}^{(i)}, \Delta\boldsymbol{\tau}^{(i)}
    =
    h_{\boldsymbol{\xi}_{\mathsf{GDN}}}
    (\hat{\eta}^{(i)}, k^{(i)}).
\end{equation}
This reduces the hypernetwork parameter complexity from $\mathcal{O}(D_{\mathsf{in}}^2)$ to $\mathcal{O}(D_{\mathsf{in}}r)$. We refer to this adaptive normalization layer as the $\mathtt{HN\text{-}G}$ layer. The same procedure is applied to IGDN layers using a separate hypernetwork $h_{\boldsymbol{\xi}_{\mathsf{IGDN}}}$, which is referred to as the $\mathtt{HN\text{-}IG}$ layer.

\begin{remark}[$\mathtt{HN\text{-}(I)G}$ layers in $\mathcal{F}_{\boldsymbol{\Theta}}$ and $\mathcal{G}_{\boldsymbol{\Phi}}$]
These layers replace the (I)GDN layers in the JSCC encoder--decoder and adaptively normalize the feature vector statistics between NN layers according to the channel conditions. For the outer encoder--decoder, the corresponding hypernetworks use $\eta^{(\mathsf{a})}$ and $k^{(\mathsf{a})}$ instead of $\hat{\eta}^{(i)}$ and $k^{(i)}$. In particular, similar to how the slimmable operation is applied to the GDN layer, the $\mathtt{HN\text{-}G}$ layer can also be integrated with the slimmable operations in \emph{Section~\ref{subsec:Implementation of the Encoder-Decoder Architecture}}. The full-dimensional adapted parameters $\boldsymbol{\tau}^{(i)} \in \mathbb{R}^{D_{\mathsf{max}}}$ and $\boldsymbol{\gamma}^{(i)} \in \mathbb{R}^{D_{\mathsf{max}}\times D_{\mathsf{max}}}$ are computed first. Then, to process a tensor with a target dimension $D_{\mathsf{target}}$, only the leading sub-vector $[\boldsymbol{\tau}^{(i)}]_{:D_{\mathsf{target}}}$ and sub-matrix $[\boldsymbol{\gamma}^{(i)}]_{:D_{\mathsf{target}},:D_{\mathsf{target}}}$ are used to perform the normalization.
\end{remark}

\subsubsection{Hypernetwork-based LN Layer}

Notably, the proposed hypernetwork-based adaptation is generally applicable to normalization layers with trainable parameters and is therefore also applied to LN layers in the DCG network. The standard LN operation, which incorporates normalization by the mean $\mu_{n}$ and variance $\sigma_{n}^2$ of $\mathbf{t}_{n}$, is given by
\begin{equation}
\label{eq:standardLN}
    \tilde{t}_{n,d,p}
    =
    \frac{t_{n,d,p} - \mu_n}{\sqrt{\sigma_n^2}}
    \cdot \alpha_d + \beta_d ,
\end{equation}
where $\boldsymbol{\alpha}, \boldsymbol{\beta} \in \mathbb{R}^{D_{\mathsf{in}}}$ are trainable scale and bias vectors. Similar to the $\mathtt{HN\text{-}(I)G}$ layer, a hypernetwork $h_{\boldsymbol{\xi}_{\mathsf{LN}}}$, which shares the same structure as $h_{\boldsymbol{\xi}_{\mathsf{GDN}}}$, generates offset vectors
\begin{equation}
    \Delta\boldsymbol{\alpha}^{(i)}, \Delta\boldsymbol{\beta}^{(i)}
    =
    h_{\boldsymbol{\xi}_{\mathsf{LN}}}
    (\hat{\eta}^{(i)}, k^{(i)}),
\end{equation}
which are added as $\boldsymbol{\alpha}^{(i)}=\boldsymbol{\alpha}+\Delta\boldsymbol{\alpha}^{(i)}$ and $\boldsymbol{\beta}^{(i)}=\boldsymbol{\beta}+\Delta\boldsymbol{\beta}^{(i)}$. Here, since LN uses only $D_{\mathsf{in}}$-dimensional tensor channel-wise scale vectors, the LoRA approach is not required. We refer to this LN-based design as the $\mathtt{HN\text{-}L}$ layer.

\begin{remark}[$\mathtt{HN\text{-}L}$ layer in $\mathcal{C}_{\boldsymbol{\Psi}}$]
This layer replaces the LN layer in $\mathtt{MLP}_{\boldsymbol{\psi}_{k^{(i)}}}$ of the $\mathcal{C}_{\boldsymbol{\Psi}}$ and adaptively normalizes intra- and inter-codeword distributions according to $\hat{\eta}^{(i)}$ and $k^{(i)}$.
\end{remark}

\vspace{-3mm}
\section{Training Strategy}
\label{sec:Training Strategy}

Directly training the entire euJSCC model over time-varying channels substantially increases the optimization difficulty and leads to unstable convergence. To address this issue, we adopt a two-phase curriculum learning strategy. In \textbf{Phase 1}, the outer encoder--decoder and the codebook-related modules are pretrained over an AWGN channel without the inner encoder--decoder to establish stable feature--codebook alignment. In \textbf{Phase 2}, the full model is finetuned over the time-varying channel with the inner modules activated.
\vspace{-3mm}
\subsection{Phase 1: Training under AWGN Channels}
\label{subsec:First Stage: Training under AWGN Channels}

Since we set $\mathbb{E}[|h_t^{(i)}|^2] = 1$, the average SNR used in the outer encoder–decoder, $\eta^{(\mathsf{a})}$, reduces to $10\log_{10}(P/\sigma^2)$, which remains constant across all blocks. Moreover, since the codebook generator also relies on $\eta^{(\mathsf{a})}$, i.e., $|h_t^{(i)}| = 1$ for $i\in[1{:}U]$, the first stage can effectively be regarded as training over a static AWGN channel. In this setting, $\mathbf{X}$ is encoded by $f_{\boldsymbol{\theta}_{\mathsf{outer}}}$, quantized using $\mathbf{C}^{(\mathsf{a})}=\mathcal{C}_{\boldsymbol{\Psi}}(\eta^{(\mathsf{a})}, k^{(\mathsf{a})})$, transmitted over the AWGN channel, and reconstructed by $g_{\boldsymbol{\phi}_{\mathsf{outer}}}$ following \textbf{Algorithm~\ref{alg:euJSCC-Forward-AWGN}}. Accordingly, $\mathbf{C}^{(\mathsf{a})}$, $\mathbf{z}^{(\mathsf{a})}$, $\mathbf{s}^{(\mathsf{a})}$, $\hat{\mathbf{s}}^{(\mathsf{a})}$, and $\hat{\mathbf{z}}^{(\mathsf{a})}$ are defined with respect to $\eta^{(\mathsf{a})}$. Since the inner encoder--decoder is omitted in \textbf{Phase 1}, the trainable parameter set is $\boldsymbol{\varphi}^{(\mathsf{i})}=\{\boldsymbol{\theta}_{\mathsf{outer}},\boldsymbol{\phi}_{\mathsf{outer}},\boldsymbol{\Psi}\}$, as summarized in \textbf{Phase 1} of \textbf{Algorithm~\ref{alg:euJSCC-Training}}.

To optimize the model across the full SNR range, the training interval is divided into $J$ subranges with boundaries $\mathbf{r}^{(\mathsf{i})}=[r^{(\mathsf{i})}_0,\dots,r^{(\mathsf{i})}_{J}]$. For each subrange $j\in[1{:}J]$, a continuous-valued $\eta^{(\mathsf{a})}$ is sampled as $\eta^{(\mathsf{a})}\sim\mathsf{Unif}[r^{(\mathsf{i})}_{j-1},r^{(\mathsf{i})}_{j})$, and the modulation order $m_k$ is determined from $\mathbf{b}$ using \eqref{eq:modulation-order-selection}. A forward pass following \textbf{Algorithm~\ref{alg:euJSCC-Forward-AWGN}} yields the loss $\mathcal{L}_j$ in~\eqref{eq:VQVAE LOSS}, and the total loss is $\mathcal{L}_{\mathsf{total}}=\sum_{j=1}^{J}\lambda^{(\mathsf{i})}_j\mathcal{L}_j$, where $\lambda^{(\mathsf{i})}_j$ compensates for differences in loss scale across SNR subranges. The parameters $\boldsymbol{\varphi}^{(\mathsf{i})}$ are then updated using Adam until the termination conditions in Section~\ref{sec:Experimental Results} are met.

\begin{algorithm}[!t]
\small
\caption{Modified euJSCC for AWGN Channels}
\label{alg:euJSCC-Forward-AWGN}
\begin{algorithmic}[1]
\State \textbf{Input: } Source image $\mathbf{X}$, average SNR $\eta^{(\mathsf{a})}$, modulation order $m_{k^{(\mathsf{a})}}$, outer encoder--decoder $({f}_{\boldsymbol{\theta}_{\mathsf{outer}}},{g}_{\boldsymbol{\phi}_{\mathsf{outer}}})$, DCG network $\mathcal{C}_{\boldsymbol{\Psi}}$
\State Encode $\mathbf{X}$ into $\mathbf{Y}_{\mathsf{mid}}$ with $\eta^{(\mathsf{a})}$ via ${f}_{\boldsymbol{\theta}_{\mathsf{outer}}}$
\State Generate $\mathbf{C}^{(\mathsf{a})}$ from $\eta^{(\mathsf{a})}$ and $k^{(\mathsf{a})}$ via $\mathcal{C}_{\boldsymbol{\Psi}}$
\State Quantize $\mathbf{Y}_{\mathsf{mid}}$ into $\mathbf{z}^{(\mathsf{a})}$ using $\mathbf{C}^{(\mathsf{a})}$
\State Modulate $\mathbf{z}^{(\mathsf{a})}$ into $\mathbf{s}^{(\mathsf{a})}$ via $m_{k^{(\mathsf{a})}}$-ary modulation
\State Transmit $\mathbf{s}^{(\mathsf{a})}$ over the channel with $\eta^{(\mathsf{a})}$ to obtain $\hat{\mathbf{s}}^{(\mathsf{a})}$
\State Demodulate $\hat{\mathbf{s}}^{(\mathsf{a})}$ into $\hat{\mathbf{z}}^{(\mathsf{a})}$ via $m_{k^{(\mathsf{a})}}$-ary demodulation
\State Dequantize $\hat{\mathbf{z}}^{(\mathsf{a})}$ into $\hat{\mathbf{Y}}_{\mathsf{mid}}$ using $\mathbf{C}^{(\mathsf{a})}$
\State Reconstruct $\hat{\mathbf{X}}$ from $\hat{\mathbf{Y}}_{\mathsf{mid}}$ with $\eta^{(\mathsf{a})}$ via ${g}_{\boldsymbol{\phi}_{\mathsf{outer}}}$
\State \textbf{Output: }$\hat{\mathbf{X}}$
\end{algorithmic}
\end{algorithm}

\begin{algorithm}[!t]
\small
\caption{euJSCC Training Algorithm}
\label{alg:euJSCC-Training}
\begin{algorithmic}[1]
\State \textbf{Input: } Training dataset $\mathcal{D}$, modulation set $\{m_k\}_{k=1}^K$, training SNR boundaries $\mathbf{r}^{(\mathsf{i})}=[r^{(\mathsf{i})}_0,\dots,r^{(\mathsf{i})}_{J}]$, $\mathbf{r}^{(\mathsf{ii})}=[r^{(\mathsf{ii})}_0,\dots,r^{(\mathsf{ii})}_{J}]$, modulation selection SNR boundaries $\mathbf{b}=[b_1,\dots,b_{K-1}]$, loss weights $\{\lambda_{j}^{(\mathsf{i})}\}_{j=1}^{J}$, $\{\lambda_{j}^{(\mathsf{ii})}\}_{j=1}^{J}$, set of data symbol lengths per transmission block $\mathcal{S}_{T}$, feedback delay set $\mathcal{S}_{\delta}$, pilot spacing $T_p$
\Statex\hrulefill

\Statex \textbf{Phase 1: Training under AWGN channels}
\State Initialize $\boldsymbol{\varphi}^{(\mathsf{i})}=\{\boldsymbol{\theta}_{\mathsf{outer}},\boldsymbol{\phi}_{\mathsf{outer}},\boldsymbol{\Psi}\}$
\While{Stopping criterion is not met}
  \State Sample $\mathbf{X} \sim \mathcal{D}$; $\mathcal{L}_{\mathsf{total}}\gets 0$\Comment{mini-batch}
  \For{$j \gets 1$ to $J$}
    \State Sample $\eta^{(\mathsf{a})} \sim \mathsf{Unif}[r^{(\mathsf{i})}_{j-1}, r^{(\mathsf{i})}_j)$
    \State Select $m_{k^{(\mathsf{a})}}$ using $\eta^{(\mathsf{a})}$ with \eqref{eq:modulation-order-selection} and $\mathbf{b}$
    \State Compute $\mathcal{L}_{j}$ as in \eqref{eq:VQVAE LOSS} via \textbf{Algorithm~\ref{alg:euJSCC-Forward-AWGN}}
    \State $\mathcal{L}_{\mathsf{total}}\gets \mathcal{L}_{\mathsf{total}}+\lambda_{j}^{(\mathsf{i})}\mathcal{L}_{j}$
  \EndFor
  \State Update $\boldsymbol{\varphi}^{(\mathsf{i})}$ with $\mathcal{L}_{\mathsf{total}}$
\EndWhile
\State \textbf{Output: } $\boldsymbol{\varphi}^{(\mathsf{i})*}=\{\boldsymbol{\theta}_{\mathsf{outer}}^*,\boldsymbol{\phi}_{\mathsf{outer}}^*,\boldsymbol{\Psi}^*\}$

\Statex\hrulefill
\Statex \textbf{Phase 2: Training under time-varying channels}
\State Initialize $\{\boldsymbol{\theta}_{\mathsf{inner}},\boldsymbol{\phi}_{\mathsf{inner}}\}$
\State Define $\boldsymbol{\varphi}^{(\mathsf{ii})}=\boldsymbol{\varphi}^{(\mathsf{i})*}\cup\{\boldsymbol{\theta}_{\mathsf{inner}},\boldsymbol{\phi}_{\mathsf{inner}}\}$
\While{Stopping criterion is not met}
  \State Sample $\mathbf{X} \sim \mathcal{D}$; $\mathcal{L}_{\mathsf{total}}\gets 0$\Comment{mini-batch}
  \For{$j \gets 1$ to $J$}
    \State Sample $T\in\mathcal{S}_{T}$ and $\delta\in\mathcal{S}_{\delta}$ uniformly
    \State Set $U=\lceil N/T\rceil$
    \State Generate $\{h_t^{(i)}\}_{t=1}^{T}$ for $i\in[1-\delta{:}U]$
    \State Estimate $\{\hat{h}_t^{(i)}\}_{t=1}^{T}$ for $i\in[1-\delta{:}U]$
    \State Compute $\eta^{(i)}$ using $\{\hat{h}_t^{(i)}\}_{t=1}^{T}$ for $i\in[1-\delta{:}U-\delta]$
    \State Quantize $\tilde{\eta}^{(i)}=\mathcal{Q}_{\mathsf{CQI}}(\eta^{(i)})$ for
    $i\in[1-\delta{:}U-\delta]$
    \State Set $\hat{\eta}_{\mathsf{Tx}}^{(i)}=\mathcal{W}\big(\tilde{\eta}^{(i-\delta)}\big)$ and $\hat{\eta}_{\mathsf{Rx}}^{(i)}=\tilde{\eta}^{(i-\delta)}$ for $i\in[1{:}U]$
    \State Select $m_{k^{(\mathsf{a})}}$ using $\eta^{(\mathsf{a})}$ with \eqref{eq:modulation-order-selection} and $\mathbf{b}$
    \State Select $\{m_{k^{(i)}}\}_{i=1}^{U}$ using $\{\hat{\eta}_{\mathsf{Tx}}^{(i)}\}_{i=1}^{U}$ with \eqref{eq:modulation-order-selection} and $\mathbf{b}$
    \State Compute $\mathcal{L}_{j}$ as in \eqref{eq:VQVAE LOSS} via \textbf{Algorithm~\ref{alg:euJSCC-Forward-FAST}}
    \State $\mathcal{L}_{\mathsf{total}}\gets \mathcal{L}_{\mathsf{total}}+\lambda_{j}^{(\mathsf{ii})}\mathcal{L}_{j}$
  \EndFor
  \State Update $\boldsymbol{\varphi}^{(\mathsf{ii})}$ with $\mathcal{L}_{\mathsf{total}}$
\EndWhile
\State \textbf{Output: } $\boldsymbol{\varphi}^{(\mathsf{ii})*}=\{\boldsymbol{\theta}_{\mathsf{outer}}^*,\boldsymbol{\phi}_{\mathsf{outer}}^*,\boldsymbol{\Psi}^*,\boldsymbol{\theta}_{\mathsf{inner}}^*,\boldsymbol{\phi}_{\mathsf{inner}}^*\}$
\end{algorithmic}
\end{algorithm}
\vspace{-3mm}
\subsection{Phase 2: Training under Time-Varying Channels}
\label{subsec:Second Stage: Training under Time-Varying Channels}

In \textbf{Phase 2}, the inner encoder--decoder pair (${f}_{\boldsymbol{\theta}_{\mathsf{inner}}}$ and ${g}_{\boldsymbol{\phi}_{\mathsf{inner}}}$) is activated, and jointly trained with the optimized parameters $\boldsymbol{\varphi}^{(\mathsf{i})*}$ from \textbf{Phase 1}. The trainable parameter set for this phase is thus initialized as $\boldsymbol{\varphi}^{(\mathsf{ii})}=\boldsymbol{\varphi}^{(\mathsf{i})*}\cup\{\boldsymbol{\theta}_{\mathsf{inner}},\boldsymbol{\phi}_{\mathsf{inner}}\}$. The whole training process for \textbf{Phase 2} is summarized in \textbf{Phase~2} of \textbf{Algorithm~\ref{alg:euJSCC-Training}}. Unlike the AWGN scenario, the channel condition varies over time, and each transmission block $i$ is explicitly considered. At each iteration, given the training SNR boundary $\mathbf{r}^{(\mathsf{ii})}=[r^{(\mathsf{ii})}_0,\dots,r^{(\mathsf{ii})}_{J}]$ for the second phase, a continuous-valued average SNR $\eta^{(\mathsf{a})}$ is randomly sampled from the continuous uniform distribution $\mathsf{Unif}[r^{(\mathsf{ii})}_{j-1}, r^{(\mathsf{ii})}_j)$, i.e., $r^{(\mathsf{ii})}_{j-1} \leq \eta^{(\mathsf{a})} < r^{(\mathsf{ii})}_j$, for $j\in[1{:}J]$. For each $j$, to allow a single euJSCC model to support different values of $T$ and $\delta$, $T\in\mathcal{S}_{T}$ and $\delta\in\mathcal{S}_{\delta}$ are randomly sampled from the corresponding discrete sets. Subsequently, the channel coefficients $\{h_t^{(i)}\}_{t=1}^{T}$ are generated and the corresponding estimates $\{\hat{h}_t^{(i)}\}_{t=1}^{T}$ are obtained for $i\in[1-\delta{:}U]$. The block-wise SNRs $\{\eta^{(i)}\}_{i=1-\delta}^{U-\delta}$ are then computed from the estimated channel coefficients and quantized as $\tilde{\eta}^{(i)}=\mathcal{Q}_{\mathsf{CQI}}(\eta^{(i)})$ for $i\in[1-\delta{:}U-\delta]$. The CQIs available for transmission block $i$ are then given by $\hat{\eta}_{\mathsf{Tx}}^{(i)}=\mathcal{W}\big(\tilde{\eta}^{(i-\delta)}\big)$ and $\hat{\eta}_{\mathsf{Rx}}^{(i)}=\tilde{\eta}^{(i-\delta)}$ for $i\in[1{:}U]$. The modulation orders $\{m_{k^{(i)}}\}_{i=1}^{U}$ and $m_{k^{(\mathsf{a})}}$ are determined using $\{\hat{\eta}_{\mathsf{Tx}}^{(i)}\}_{i=1}^{U}$ and $\eta^{(\mathsf{a})}$, respectively, with the boundary set $\mathbf{b}$.

The loss $\mathcal{L}_j$ for each SNR subrange is computed according to \eqref{eq:VQVAE LOSS} following \textbf{Algorithm~\ref{alg:euJSCC-Forward-FAST}}. The total loss for the second phase is obtained by aggregating these subrange losses $\mathcal{L}_{\mathsf{total}} = \sum_{j=1}^{J} \lambda^{(\mathsf{ii})}_j\, \mathcal{L}_j$, where $\lambda^{(\mathsf{ii})}_j$ balances the contributions of different SNR subranges. Finally, the parameter set $\boldsymbol{\varphi}^{(\mathsf{ii})}=\{{\boldsymbol{\theta}_{\mathsf{outer}}}, {\boldsymbol{\phi}_{\mathsf{outer}}}, \boldsymbol{\Psi}, {\boldsymbol{\theta}_{\mathsf{inner}}}, {\boldsymbol{\phi}_{\mathsf{inner}}}\}$ is iteratively updated using the Adam optimizer. Training continues until the stopping criterion specified in Section~\ref{sec:Experimental Results} is satisfied.

\begin{table*}[t]
\caption{NN structure of the euJSCC encoder and decoder, and hyperparameter and channel settings for experiments. The notation $\{\cdot\}_{k=1}^{K}$ denotes $K$ parallel layers.}
\label{tab:euJSCC_structure}
\label{tab:hyperparameters}
\centering

\begin{minipage}[t]{0.69\textwidth}
\centering
\resizebox{\linewidth}{!}{%
\begin{tabular}{@{}lll@{}}
\toprule
\multicolumn{1}{c}{\textbf{Module}} & 
\multicolumn{1}{c}{\textbf{Layers}} & 
\multicolumn{1}{c}{\textbf{Output size}} \\
\midrule

\multirow{6}{*}{\textbf{Outer encoder}} 
 & $\mathtt{Conv2D}(C, c_1, 4, 2, 1)$ + $\mathtt{HN\text{-}G}(c_1)$ + $\mathtt{PReLU}$ 
 & $c_1\times H/2\times W/2$ \\

 & $\mathtt{Conv2D}(c_1, c_2, 4, 2, 1)$ + $\mathtt{HN\text{-}G}(c_2)$ + $\mathtt{PReLU}$ 
 & $c_2\times H/4\times W/4$ \\

 & $\mathtt{Res}(c_2)$ + $\mathtt{PReLU}$ 
 & $c_2\times H/4\times W/4$ \\

 & $\mathtt{Conv2D}(c_2, c_1, 5, 1, 2)$ + $\mathtt{HN\text{-}G}(c_1)$ + $\mathtt{PReLU}$ 
 & $c_1\times H/4\times W/4$ \\

 & $\mathtt{Res}(c_1)$ + $\mathtt{PReLU}$ 
 & $c_1\times H/4\times W/4$ \\

 & $\mathtt{Conv2D}(c_1, D_K, 5, 1, 2)$ + $\mathtt{HN\text{-}G}(D_K)$ + $\mathtt{Reshape}$ 
 & $D_{k^{(\mathsf{a})}}\times HW/16$ \\

\midrule

\textbf{Inner encoder} 
 & $\{\mathtt{Conv1D}(D_k, D_K, 5, 1, 2)\}_{k=1}^{K}$ + $\{\mathtt{HN\text{-}G}(D_K)\}_{k=1}^{K}$ 
 & $D_{k^{(i)}}\times HW/16$ \\

\midrule

\textbf{Inner decoder} 
 & $\{\mathtt{Conv1D}(D_k, D_K, 5, 1, 2)\}_{k=1}^{K}$ + $\{\mathtt{HN\text{-}IG}(D_k)\}_{k=1}^{K}$ 
 & $D_{k^{(\mathsf{a})}}\times HW/16$ \\

\midrule

\multirow{6}{*}{\textbf{Outer decoder}} 
 & $\mathtt{Reshape}$ + $\mathtt{T\text{-}Conv2D}(D_K, c_1, 5, 1, 2)$ + $\mathtt{HN\text{-}IG}(c_1)$ + $\mathtt{PReLU}$ 
 & $c_1\times H/4\times W/4$ \\

 & $\mathtt{T\text{-}Res}(c_1)$ + $\mathtt{PReLU}$ 
 & $c_1\times H/4\times W/4$ \\

 & $\mathtt{T\text{-}Conv2D}(c_1, c_2, 5, 1, 2)$ + $\mathtt{HN\text{-}IG}(c_2)$ + $\mathtt{PReLU}$ 
 & $c_2\times H/4\times W/4$ \\

 & $\mathtt{T\text{-}Res}(c_2)$ + $\mathtt{PReLU}$ 
 & $c_2\times H/4\times W/4$ \\

 & $\mathtt{T\text{-}Conv2D}(c_2, c_1, 4, 2, 1)$ + $\mathtt{HN\text{-}IG}(c_1)$ + $\mathtt{PReLU}$ 
 & $c_1\times H/2\times W/2$ \\

 & $\mathtt{T\text{-}Conv2D}(c_1, C, 4, 2, 1)$ + $\mathtt{HN\text{-}IG}(C)$ + $\mathtt{Reshape}$ 
 & $C\times H\times W$ \\

\midrule

\multirow{2}{*}{$\mathtt{Res}(c)$} 
 & $\mathtt{Conv2D}(c, c, 3, 1, 1)$ + $\mathtt{HN\text{-}G}(c)$ + $\mathtt{PReLU}$ 
 & $c\times h\times w$ \\

 & $\mathtt{Conv2D}(c, c, 1, 1, 0)$ + $\mathtt{HN\text{-}G}(c)$ 
 & $c\times h\times w$ \\

\midrule

\multirow{2}{*}{$\mathtt{T\text{-}Res}(c)$} 
 & $\mathtt{T\text{-}Conv2D}(c, c, 1, 1, 0)$ + $\mathtt{HN\text{-}IG}(c)$ + $\mathtt{PReLU}$ 
 & $c\times h\times w$ \\

 & $\mathtt{T\text{-}Conv2D}(c, c, 3, 1, 1)$ + $\mathtt{HN\text{-}IG}(c)$ 
 & $c\times h\times w$ \\

\bottomrule
\end{tabular}
}
\end{minipage}
\hfill
\begin{minipage}[t]{0.28\textwidth}
\centering

\resizebox{\linewidth}{!}{%
\begin{tabular}{@{}ll@{}}
\toprule
\multicolumn{2}{c}{\textbf{Hyperparameters}} \\
\midrule

$J$ & $5$ \\

$\mathbf{r}^{(\mathsf{i})}$ 
& $-5,~5,~12,~20,~26,~35$ \\

$\mathbf{r}^{(\mathsf{ii})}$ 
& $3,~8,~13,~18,~23,~27$ \\

$\mathbf{b}$ 
& $5,~12,~20,~26$ \\

$\{\lambda_{j}^{(\mathsf{i})}\}_{j=1}^{J}$ 
& $1,~2,~3,~6,~12$ \\

$\{\lambda_{j}^{(\mathsf{ii})}\}_{j=1}^{J}$ 
& $1,~2,~3,~6,~12$ \\

$\{\alpha_k\}_{k=1}^{K}$ 
& $3,~2,~1,~0.7,~0.5$ \\

\midrule
\multicolumn{2}{c}{\textbf{Channel Parameters}} \\
\midrule

Channel model 
& CDL-A \\

$\mathcal{S}_{T}$ 
& $256,~512,~1024$ \\

$\mathcal{S}_{\delta}$ 
& $0,~1,~2$ \\

Symbol duration 
& $1~\mu\mathrm{s}$ \\

Carrier frequency 
& $3.5$ GHz \\

User velocity $V$ 
& $60$ km/h \\

$B_{\mathsf{CQI}}$ 
& $4$ \\

\bottomrule
\end{tabular}
}
\end{minipage}

\vspace{-4mm}
\end{table*}

\vspace{-1mm}
\section{Experimental Results}\label{sec:Experimental Results}

The proposed euJSCC model is trained on the DIV2K training dataset using randomly cropped patches of size $256 \times 256$. The dataset is divided into training and validation subsets in a 9:1 ratio, with validation performed after every epoch. If no improvement is observed over 15 consecutive validations, the learning rate is reduced by a factor of $0.6$, and training is terminated after 100 consecutive validations without improvement. Performance is evaluated on the Kodak dataset with images of size $512 \times 768$ and the CLIC2021 test set, which consists of high-resolution images of approximately 2K resolution. The channel bandwidth ratio (CBR), defined as the ratio between the number of transmitted data symbols and the dimensionality of the source data, is set to $\frac{N}{CHW}=\frac{1}{48}$.

The network architecture of the euJSCC encoder--decoder is summarized in Table~\ref{tab:euJSCC_structure}. Here, $\mathtt{Conv2D}(c_{\mathsf{in}},c_{\mathsf{out}},f,s,p)$ and $\mathtt{T\text{-}Conv2D}(c_{\mathsf{in}},c_{\mathsf{out}},f,s,p)$ denote 2D convolutional and transposed convolutional layers with input and output channel dimensions $c_{\mathsf{in}}$ and $c_{\mathsf{out}}$, kernel size $f$, stride $s$, and padding $p$, while $\mathtt{Conv1D}$ follows the same notation. $\mathtt{Res}(c)$ and $\mathtt{T\text{-}Res}(c)$ denote residual and transposed residual blocks with channel dimension $c$, and $\mathtt{HN\text{-}G}(c)$ and $\mathtt{HN\text{-}IG}(c)$ denote the proposed hypernetwork-based GDN and inverse GDN layers, respectively.

For a fair comparison, all baselines use the same time-varying channel, pilot-based channel estimation, CQI feedback procedure, and feedback-delay settings, unless otherwise specified. The baselines employed for comparison are listed below:

\begin{itemize}
    \item \textbf{euJSCC:} The full euJSCC model is trained through both phases of \textbf{Algorithm~\ref{alg:euJSCC-Training}} and evaluated under the considered time-varying channels.

    \item \textbf{uJSCC:} Since the original uJSCC~\cite{huh2025universal} is not designed for time-varying channels, we extend it for a fair comparison by inserting an inner encoder and decoder with the same architecture as those in euJSCC. Their normalization layers are implemented as switchable (inverse) GDN layers comprising $K^2$ modules, each selected according to $(k^{(\mathsf{a})},k^{(i)})$.

    \item \textbf{MDJCM:} MDJCM~\cite{zhang2024analog} is a channel-adaptive SQ-JSCC model with SNR- and modulation-conditioned feature adaptation and entropy-based adjustment of feature vector lengths. For a fair comparison, we follow its original three-phase training procedure, replace the Swin Transformer backbone with a CNN-based architecture, and extend it to the considered time-varying channels by applying its adaptation modules to each transmission block using the available CQI feedback and corresponding modulation order. The side information indicating the feature lengths is on the order of $10^3$--$10^4$ bits per image under the considered settings and should be reliably delivered to the receiver. Accordingly, it is transmitted using NR LDPC with adaptive MCS selection targeting a block error rate (BLER) of 0.1 and Type-I HARQ before feature vector transmission.

    \item \textbf{SSCC:} As separate source and channel coding (SSCC) baselines, we employ JPEG2000 and NTC~\cite{balle2020nonlinear} for source coding and LDPC codes for channel coding, with the encoded bitstream transmitted using the same adaptive MCS selection and Type-I HARQ procedure as the MDJCM side information. These two baselines are referred to as SSCC-JPEG2000 and SSCC-NTC, respectively.
\end{itemize}

\textbf{Phase 1} employs a static AWGN channel, whereas \textbf{Phase 2} employs time-varying flat fading channel realizations directly sampled from the 3GPP clustered delay line A (CDL-A) model without resampling. The learning rate and batch size are set to $10^{-4}$ and $4$, respectively. The system supports $K=5$ modulation orders, namely BPSK, 4-QAM, 16-QAM, 64-QAM, and 256-QAM, with $m_k\in\{2,4,16,64,256\}$ and corresponding feature vector channel dimensions $D_k\in\{4,8,16,24,32\}$ empirically selected to improve feature vector expressiveness while limiting quantization distortion~\cite{huh2025universal}. For all $\mathtt{HN\text{-}(I)G}$ layers, the low rank $r$ is set to $4$, and $\beta_k$ is set to $\alpha_k/4$ for all $k\in[1{:}K]$. $T_p$ is set to $10$, resulting in the pilot overhead of $9.38\%$. The remaining hyperparameters and channel settings are summarized in Table~\ref{tab:hyperparameters}.

Following the setting in~\cite{huh2025universal}, $\mathbf{b}$ is selected to target a symbol error rate (SER) of approximately $0.1$. The selected boundaries are representative rather than intrinsic to euJSCC, and the impact of different boundary settings is evaluated in Section~\ref{subsec:Performance Under Various Channel and System Conditions}. Image transmission performance is evaluated using peak signal-to-noise ratio (PSNR) and multi-scale structural similarity (MS-SSIM), with MS-SSIM reported in dB as $\text{MS-SSIM (dB)}=-10\log_{10}(1-\text{MS-SSIM})$.

\vspace{-4mm}
\subsection{Complexity Analysis}

\begin{table}[!t]
\caption{Number of parameters across models. For euJSCC, the percentage indicates the portion of parameters in each listed module relative to the full euJSCC model.}
\vspace{-1mm}
\label{tab:num_params}
\centering
\begin{tabular}{@{}lrr@{}}
\toprule
\multicolumn{1}{c}{\textbf{Model}} & 
\multicolumn{1}{c}{\textbf{$\#$ of Params.}} &
\multicolumn{1}{c}{\textbf{Portion (\%)}} \\
\midrule
euJSCC & 9{,}759{,}356 & -- \\
\quad $\mathtt{HN\text{-}(I)G}$ & 1{,}117{,}902 & 11.5 \\
\quad\quad $h_{\boldsymbol{\xi}_{\mathsf{(I)GDN}}}$ for $\mathtt{HN\text{-}(I)G}$ & 197{,}312 & 2.0 \\
\quad\quad ($h_{\boldsymbol{\xi}_{\mathsf{(I)GDN}}}$ w/o LoRA) & 5{,}532{,}936 & -- \\
\quad $\mathcal{C}_{\boldsymbol{\Psi}}$ & 158{,}429 & 1.6 \\
\quad ${f}_{\boldsymbol{\theta}_{\mathsf{inner}}}$, ${g}_{\boldsymbol{\phi}_{\mathsf{inner}}}$ & 47{,}566 & 0.5 \\
\midrule
uJSCC~\cite{huh2025universal} & 13{,}100{,}112 & -- \\
MDJCM~\cite{zhang2024analog} & 11{,}006{,}018 & -- \\
\bottomrule
\end{tabular}
\vspace{-3mm}
\end{table}

\begin{table}[!t]
\caption{FLOPs and inference time of euJSCC at an average SNR of $8$~dB. The values separated by ``/'' correspond to the transmitter- and receiver-side operations, respectively. The inner modules, VQ, VDQ, and DCG are reported per block. Indented rows are included in the corresponding preceding entries.}
\vspace{-1mm}
\label{tab:complexity}
\centering
\setlength{\tabcolsep}{4pt}
\begin{tabular}{@{}lcc@{}}
\toprule
\textbf{Module}
& \textbf{FLOPs}
& \textbf{Inference Time (ms)} \\
\midrule

Full euJSCC
& $77.8935$ G
& $2.0427$ \\
\midrule

\quad $f_{\boldsymbol{\theta}_{\mathsf{outer}}} /
g_{\boldsymbol{\phi}_{\mathsf{outer}}}$
& $38.9443/38.9449$ G
& $0.8248/0.8531$ \\

\quad\quad $h_{\boldsymbol{\xi}_{\mathsf{(I)GDN}}}$
& $3.842/3.832$ M
& $0.0276/0.0274$ \\
\midrule

\quad $f_{\boldsymbol{\theta}_{\mathsf{inner}}} /
g_{\boldsymbol{\phi}_{\mathsf{inner}}}$
& $0.245/0.272$ M
& $0.0183/0.0179$ \\

\quad\quad $h_{\boldsymbol{\xi}_{\mathsf{(I)GDN}}}$
& $0.0111/0.00123$ M
& $0.0117/0.0108$ \\
\midrule

\quad VQ / VDQ
& $0.0151/0.00612$ M
& $0.0330/0.0209$ \\

\quad\quad $\mathcal{C}_{\boldsymbol{\Psi}}$
& $0.000741/0.000741$ M
& $0.0142/0.0140$ \\
\bottomrule
\end{tabular}
\vspace{-3mm}
\end{table}

As shown in Table~\ref{tab:num_params}, the hypernetworks $h_{\boldsymbol{\xi}_{\mathsf{(I)GDN}}}$ and the DCG network $\mathcal{C}_{\boldsymbol{\Psi}}$ account for only about $3.6\%$ of the total parameters in euJSCC. Without LoRA, $h_{\boldsymbol{\xi}_{\mathsf{(I)GDN}}}$ requires over $28\times$ more parameters than its LoRA-based implementation. Moreover, $f_{\boldsymbol{\theta}_{\mathsf{inner}}}$ and $g_{\boldsymbol{\phi}_{\mathsf{inner}}}$ account for only about $0.5\%$, demonstrating that the proposed adaptive design incurs minimal memory overhead. Table~\ref{tab:complexity} reports the computational complexity and inference time of euJSCC, averaged over 200 runs using two NVIDIA GeForce RTX 4090 GPUs for the transmitter- and receiver-side processing, respectively, with a batch size of 1 using $256\times256$ input images at an average SNR of $8$~dB. $f_{\boldsymbol{\theta}_{\mathsf{outer}}}$, $g_{\boldsymbol{\phi}_{\mathsf{outer}}}$, and their $h_{\boldsymbol{\xi}_{\mathsf{(I)GDN}}}$ are executed once per image, whereas $f_{\boldsymbol{\theta}_{\mathsf{inner}}}$, $g_{\boldsymbol{\phi}_{\mathsf{inner}}}$, their $h_{\boldsymbol{\xi}_{\mathsf{(I)GDN}}}$, VQ, vector dequantization (VDQ), and $\mathcal{C}_{\boldsymbol{\Psi}}$ are executed for each transmission block. Since the layer-wise $h_{\boldsymbol{\xi}_{\mathsf{(I)GDN}}}$ and $\mathcal{C}_{\boldsymbol{\Psi}}$ depend only on the available channel information and corresponding modulation order, their computations do not depend on one another and can be performed in parallel, with the transmitter- and receiver-side operations executed on separate GPUs. In addition, receiver-side VDQ and inner decoding can begin as each transmission block is received while the transmitter continues processing the remaining transmission blocks, allowing the transmitter- and receiver-side computations to proceed concurrently. The reported inference time reflects this concurrent implementation. These block-wise operations require only approximately $0.54$~MFLOPs per block. For $T=512$, $U=8$ transmission blocks require approximately $4.3$~MFLOPs in total, corresponding to only about $0.0055\%$ of the full euJSCC FLOPs. Their short inference times further demonstrate that repeated channel adaptation incurs minimal computational overhead.
\vspace{-3mm}

\begin{figure}[!t]
  \centering
  \includegraphics[width=0.9\linewidth]{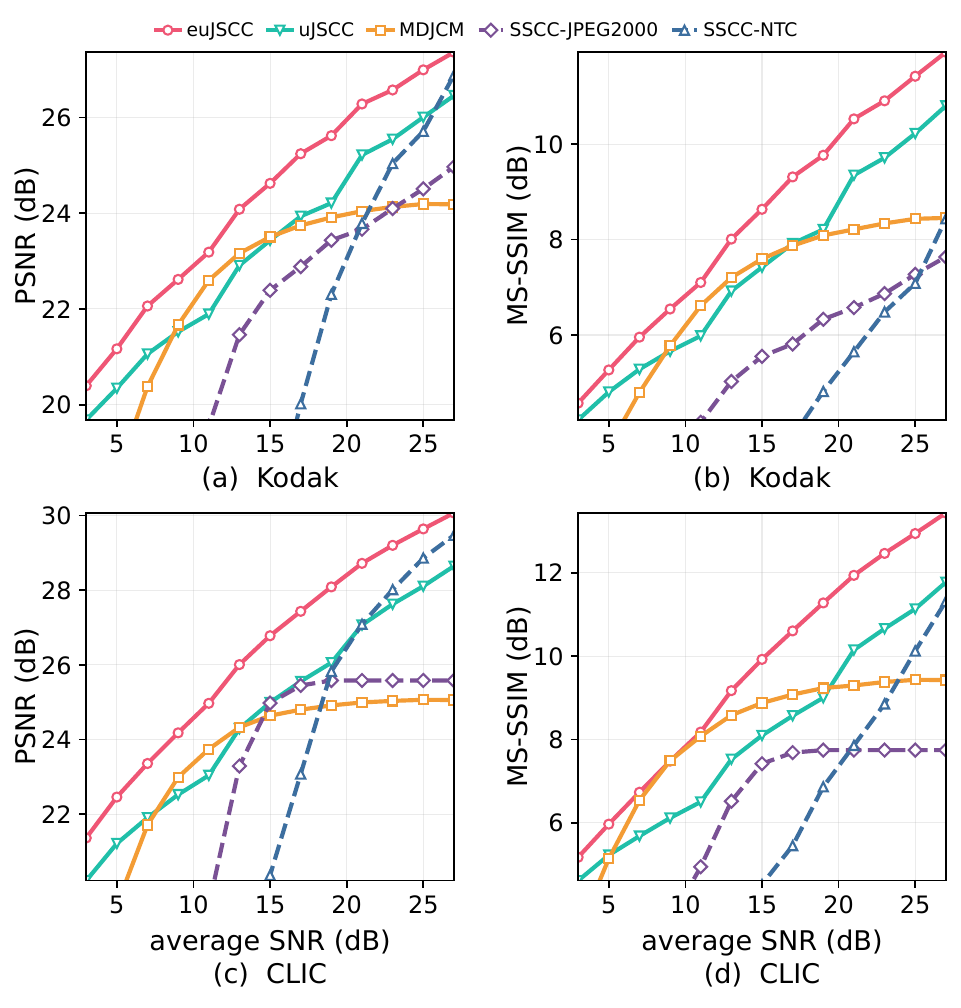}
  \vspace{-5mm}
  \caption{Performance comparison according to average SNR under time-varying channels on the Kodak and CLIC2021 test sets.}
  \label{fig:results_block}
  \vspace{-3mm}
\end{figure}

\begin{figure}[!t]
  \centering
  \includegraphics[width=0.9\linewidth]{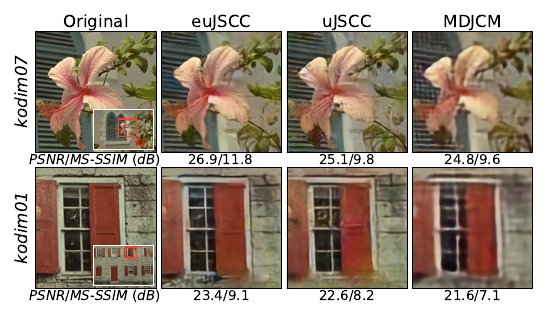}
  \vspace{-5mm}
  \caption{Visualization example of the reconstructed images on Kodak with average SNR of $20$~dB.}
  \label{fig:visual_results}
  \vspace{-5mm}
\end{figure}

\subsection{Performance Analysis}\label{subsec:Performance Analysis}

\begin{figure*}[!ht]
  \centering
  \includegraphics[width=0.9\linewidth]{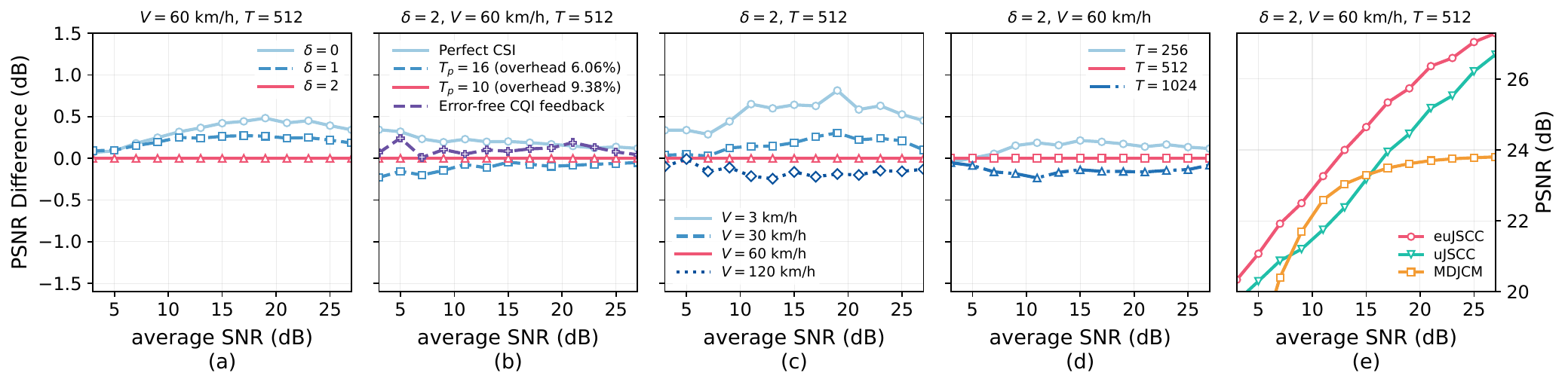}
  \vspace{-5mm}
  \caption{Reconstruction performance under various communication scenarios. (a)--(d) report the PSNR difference relative to the corresponding reference setting: (a) feedback delay $\delta$ with $\delta=2$ as the reference, (b) pilot overhead and CQI feedback errors with $T_p=10$ as the reference, (c) user velocity $V$ with $V=60$~km/h as the reference, and (d) data symbol length per transmission block $T$ with $T=512$ as the reference. Panel (e) compares the reconstruction performance under a different modulation-selection boundary set $\mathbf{b}$.}
  \label{fig:channel_analysis}
  \vspace{-5mm}
\end{figure*}

We evaluate the performance under the considered time-varying channels by averaging the results over 50 independent channel realizations for each input image and over the test dataset. Fig.~\ref{fig:results_block} presents the PSNR and MS-SSIM results of euJSCC and the baselines on the Kodak and CLIC2021 datasets with $V=60$~km/h, $\delta=2$, and $T=512$. Across both datasets, euJSCC consistently outperforms uJSCC in both metrics. Since both schemes use the same modulation orders and corresponding feature vector channel dimensions and codebook sizes, this performance improvement demonstrates the effectiveness of the additional fine-grained channel adaptation in euJSCC. Although MDJCM performs entropy-based feature vector length adaptation, its side information must be reliably transmitted to the receiver. At lower average SNRs, reliable side-information transmission requires a larger portion of the transmission budget, thereby reducing the number of feature vectors that can be transmitted and causing more severe performance degradation. At higher average SNRs, MDJCM saturates, whereas euJSCC continues to improve through fine-grained channel adaptation and the use of higher-dimensional feature vectors and larger VQ codebooks with higher-order modulations. For the SSCC baselines, SSCC-NTC achieves performance close to euJSCC at high average SNRs, whereas both SSCC baselines degrade substantially at moderate-to-low SNRs because error-free recovery of the source-coded bitstream requires an increasing number of transmitted symbols as the channel condition deteriorates. Fig.~\ref{fig:visual_results} further presents reconstructed images on the Kodak dataset at an average SNR of $20$~dB under the same channel realization. Consistent with the quantitative results, euJSCC better preserves fine image details and local structures than uJSCC and MDJCM. Overall, these results demonstrate that euJSCC maintains high reconstruction performance under time-varying channels.

\vspace{-3mm}
\subsection{Performance Under Various Communication Scenarios}\label{subsec:Performance Under Various Channel and System Conditions}

Fig.~\ref{fig:channel_analysis} evaluates euJSCC under various channel and system conditions, where Figs.~\ref{fig:channel_analysis}(a)--(d) report the PSNR difference relative to the corresponding reference setting. Unless otherwise specified, $V=60$~km/h, $\delta=2$, $T=512$, and $T_p=10$ are used. Fig.~\ref{fig:channel_analysis}(a) shows that reducing the feedback delay $\delta$ improves PSNR by providing a more recent CQI that better reflects the current channel condition. Fig.~\ref{fig:channel_analysis}(b) evaluates the effects of pilot overhead and CQI feedback errors. Compared with perfect channel state information (CSI) at the receiver, pilot-based channel estimation degrades performance by providing less accurate channel information for both equalization and CQI-based adaptation, and the degradation increases as the pilot overhead is reduced. The impact of the CQI feedback error is also evaluated by comparing with error-free CQI feedback, and the resulting performance degradation remains modest. Fig.~\ref{fig:channel_analysis}(c) shows that a lower user velocity $V$ improves PSNR because slower channel variation makes the available CQI more representative of the current channel condition, whereas a higher $V$ causes degradation due to more rapid variation. Fig.~\ref{fig:channel_analysis}(d) shows that a smaller $T$ improves reconstruction performance through more frequent CQI feedback and channel adaptation. Finally, Fig.~\ref{fig:channel_analysis}(e) shows that euJSCC continues to outperform uJSCC and MDJCM with $\mathbf{b}=\{3,10,14,24\}$, demonstrating that its performance advantage is not limited to a particular choice of modulation-selection boundaries.

\vspace{-4mm}
\subsection{Codebook Analysis}

\begin{figure}[!t]
  \centering
  \includegraphics[width=0.85\linewidth]{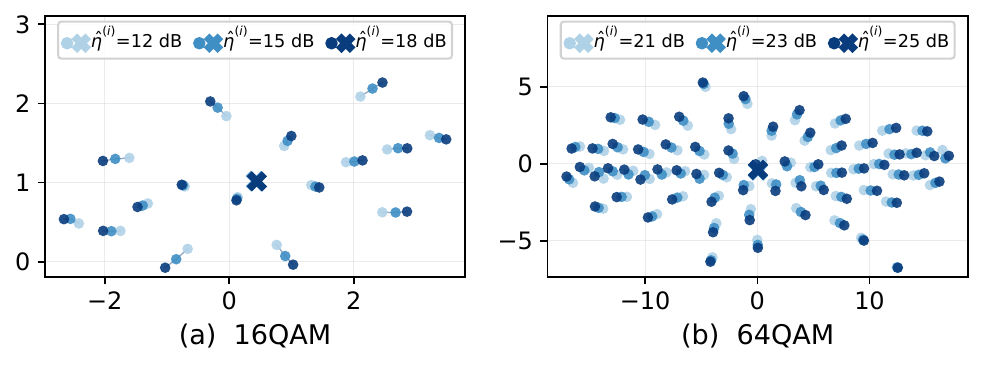}
  \vspace{-4mm}
  \caption{A t-SNE visualization of the codebooks generated from the DCG network for (a) 16-QAM and (b) 64-QAM for different input CQI values. Cross markers represent the t-SNE visualization of the mean codewords for each CQI value. The trajectory for each codeword indicates how the codeword shifts as the CQI varies.}
  \label{fig:codebook_tsne}
  \vspace{-5mm}
\end{figure}

To examine how $\mathcal{C}_{\boldsymbol{\Psi}}$ generates euJSCC codebooks across modulation orders and CQI values, Figs.~\ref{fig:codebook_tsne}(a) and (b) visualize their t-SNE distributions for 16-QAM and 64-QAM, respectively. t-SNE approximately preserves pairwise codeword similarities in two dimensions, where proximity indicates similarity and the overall spread reflects expressiveness. For both modulations, the mean codeword locations remain similar across CQI values, while the codewords are more tightly clustered at lower CQIs and more dispersed at higher CQIs. This indicates that euJSCC favors similar codewords under poorer channel conditions for robustness against channel noise and increases codeword diversity under better conditions for greater expressiveness, adaptively balancing robustness and expressiveness according to the available CQI.

\vspace{-3mm}
\subsection{Evaluation on Downstream Task}

To further evaluate euJSCC on a downstream semantic task, we consider semantic segmentation, which assigns one of the predefined semantic classes to each pixel. Experiments are conducted on CamVid with $C_{\mathsf{seg}}=11$ classes and $480\times360$ images, using the training and test sets for training and evaluation, respectively. Let $\mathbf{L}\in\{0,1\}^{C_{\mathsf{seg}}\times H\times W}$ denote the one-hot ground-truth segmentation map. All considered JSCC schemes are modified by replacing the decoder output with pixel-wise class probabilities and the image reconstruction loss with pixel-wise cross-entropy. Accordingly, for euJSCC, the outer decoder in~\eqref{eq:outer_decoding} outputs $\hat{\mathbf{P}}\in[0,1]^{C_{\mathsf{seg}}\times H\times W}$ instead of $\hat{\mathbf{X}}$, and $\mathsf{MSE}(\hat{\mathbf{X}},\mathbf{X})$ in~\eqref{eq:VQVAE LOSS} is replaced with cross-entropy loss $\mathsf{CE}(\hat{\mathbf{P}},\mathbf{L})$, while the codebook and commitment loss terms remain unchanged. The training procedure and communication conditions are otherwise the same as those used in Section~\ref{subsec:Performance Analysis}. As reported in Table~\ref{tab:semantic_segmentation}, euJSCC achieves the highest pixel accuracy and mean intersection-over-union (mIoU) at average SNRs of $3$, $9$, and $15$~dB, demonstrating its effectiveness for semantic segmentation.

\begin{table}[!t]
\caption{Semantic segmentation under different average SNRs on CamVid dataset. Each entry reports Pixel Accuracy (mIoU).}
\vspace{-1mm}
\label{tab:semantic_segmentation}
\centering
\setlength{\tabcolsep}{5pt}
\begin{tabular}{@{}cccc@{}}
\toprule
\multirow{2}{*}{\textbf{Model}} &
\multicolumn{3}{c}{\textbf{Average SNR}} \\
\cmidrule(lr){2-4}
&
\textbf{3 dB} &
\textbf{9 dB} &
\textbf{15 dB} \\
\midrule
euJSCC &
\textbf{0.813 (0.360)} &
\textbf{0.852 (0.454)} &
\textbf{0.875 (0.494)} \\

uJSCC &
0.717 (0.296) &
0.813 (0.414) &
0.857 (0.471) \\

MDJCM &
0.634 (0.244) &
0.702 (0.318) &
0.723 (0.343) \\
\bottomrule
\end{tabular}
\vspace{-5mm}
\end{table}

\vspace{-1mm}
\section{Conclusion}

In this paper, we proposed an euJSCC framework that enables semantic communication to adapt to both modulation orders and fine-grained channel conditions within a unified model. By integrating a hypernetwork-based normalization layer and a DCG network, euJSCC adaptively processes feature representations and generates VQ codebooks according to the available CQI and modulation order while maintaining low parameter and computational overhead. Moreover, the inner--outer encoder--decoder architecture, combined with a two-phase training strategy, enables effective adaptation to long-term and short-term channel conditions under time-varying channels with periodic and delayed CQI feedback. Extensive experiments demonstrate that euJSCC consistently outperforms existing channel-adaptive digital JSCC schemes in terms of PSNR and MS-SSIM across a wide range of SNRs while maintaining parameter and computational efficiency. Additional evaluations under practical channel and feedback conditions and on a downstream semantic task further validate its effectiveness across diverse communication scenarios.

\ifCLASSOPTIONcaptionsoff
  \newpage
\fi

\bibliographystyle{IEEEtran}
\bibliography{IEEEabrv,reference}

\begin{thebibliography}{10}
\providecommand{\url}[1]{#1}
\csname url@samestyle\endcsname
\providecommand{\newblock}{\relax}
\providecommand{\bibinfo}[2]{#2}
\providecommand{\BIBentrySTDinterwordspacing}{\spaceskip=0pt\relax}
\providecommand{\BIBentryALTinterwordstretchfactor}{4}
\providecommand{\BIBentryALTinterwordspacing}{\spaceskip=\fontdimen2\font plus
\BIBentryALTinterwordstretchfactor\fontdimen3\font minus \fontdimen4\font\relax}
\providecommand{\BIBforeignlanguage}[2]{{%
\expandafter\ifx\csname l@#1\endcsname\relax
\typeout{** WARNING: IEEEtran.bst: No hyphenation pattern has been}%
\typeout{** loaded for the language `#1'. Using the pattern for}%
\typeout{** the default language instead.}%
\else
\language=\csname l@#1\endcsname
\fi
#2}}
\providecommand{\BIBdecl}{\relax}
\BIBdecl

\bibitem{shao2024theory}
Y.~Shao, Q.~Cao, and D.~G{\"u}nd{\"u}z, ``A theory of semantic communication,'' \emph{IEEE Transactions on Mobile Computing}, vol.~23, no.~12, pp. 12\,211--12\,228, 2024.

\bibitem{shi2023task}
Y.~Shi, Y.~Zhou, D.~Wen, Y.~Wu, C.~Jiang, and K.~B. Letaief, ``Task-oriented communications for {6G}: Vision, principles, and technologies,'' \emph{IEEE Wireless Communications}, vol.~30, no.~3, pp. 78--85, 2023.

\bibitem{seo2025contextual}
H.~Seo, Y.~Huh, H.~Song, W.~Choi, and M.~Bennis, ``Contextual reasoning-based semantics-native communication,'' in \emph{Foundations of Semantic Communication Networks}.\hskip 1em plus 0.5em minus 0.4em\relax Hoboken, NJ, USA: Wiley, 2025, ch.~9, pp. 181--200.

\bibitem{huh2025feature}
Y.~Huh, B.~Kim, and W.~Choi, ``Feature reconstruction aided federated learning for image semantic communication,'' in \emph{IEEE Global Communications Conference (GLOBECOM)}, 2025, pp. 4669--4674.

\bibitem{huh2026differential}
------, ``Differential privacy in feature reconstruction aided federated learning for agent's semantic communication model update,'' in \emph{IEEE Global Communications Conference (GLOBECOM)}, 2026, pp. 1--6.

\bibitem{farsad2018deep}
N.~Farsad, M.~Rao, and A.~Goldsmith, ``Deep learning for joint source-channel coding of text,'' in \emph{IEEE International Conference on Acoustics, Speech and Signal Processing (ICASSP)}, 2018, pp. 2326--2330.

\bibitem{zhang2025toward}
G.~Zhang, K.~Zhou, Y.~Cai, Q.~Hu, and G.~Yu, ``Toward compatible semantic communication: A perspective on digital coding and modulation,'' \emph{IEEE Communications Magazine}, vol.~63, no.~12, pp. 156--162, 2025.

\bibitem{guo2024digital}
L.~Guo, W.~Chen, Y.~Sun, and B.~Ai, ``{Digital-SC}: Digital semantic communication with adaptive network split and learned non-linear quantization,'' \emph{IEEE Transactions on Cognitive Communications and Networking}, vol.~11, no.~4, pp. 2499--2511, 2025.

\bibitem{tung2022deepjscc}
T.-Y. Tung, D.~B. Kurka, M.~Jankowski, and D.~G{\"u}nd{\"u}z, ``{DeepJSCC-Q}: Constellation constrained deep joint source-channel coding,'' \emph{IEEE Journal on Selected Areas in Information Theory}, vol.~3, no.~4, pp. 720--731, 2022.

\bibitem{zhang2024analog}
G.~Zhang, P.~Yang, Y.~Cai, Q.~Hu, and G.~Yu, ``From analog to digital: Multi-order digital joint coding-modulation for semantic communication,'' \emph{IEEE Transactions on Communications}, vol.~73, no.~6, pp. 4257--4271, 2025.

\bibitem{xie2023robust}
S.~Xie, S.~Ma, M.~Ding, Y.~Shi, M.~Tang, and Y.~Wu, ``Robust information bottleneck for task-oriented communication with digital modulation,'' \emph{IEEE Journal on Selected Areas in Communications}, vol.~41, no.~8, pp. 2577--2591, 2023.

\bibitem{zheng2025deep}
W.~Zheng, H.~Liang, C.~Dong, and X.~Xu, ``Deep joint source-channel coding based on feedback-driven codebook optimization,'' in \emph{IEEE Wireless Communications and Networking Conference (WCNC)}, 2025, pp. 1--6.

\bibitem{huh2025universal}
Y.~Huh, H.~Seo, and W.~Choi, ``Universal joint source-channel coding for modulation-agnostic semantic communication,'' \emph{IEEE Journal on Selected Areas in Communications}, vol.~43, no.~7, pp. 2560--2574, 2025.

\bibitem{huh2026markov}
Y.~Huh, J.~Kang, and W.~Choi, ``Markov-enforced discrete diffusion model for digital semantic symbol error correction,'' \emph{IEEE Transactions on Communications}, 2026, to be published.

\bibitem{lee2023wireless}
J.~Lee, H.~Lee, and W.~Choi, ``Wireless channel adaptive {DNN} split inference for resource-constrained edge devices,'' \emph{IEEE Communications Letters}, vol.~27, no.~6, pp. 1520--1524, 2023.

\bibitem{balle2020nonlinear}
J.~Ball{\'e}, P.~A. Chou, D.~Minnen, S.~Singh, N.~Johnston, E.~Agustsson, S.~J. Hwang, and G.~Toderici, ``Nonlinear transform coding,'' \emph{IEEE Journal of Selected Topics in Signal Processing}, vol.~15, no.~2, pp. 339--353, 2021.

\bibitem{hu2018squeeze}
J.~Hu, L.~Shen, and G.~Sun, ``Squeeze-and-excitation networks,'' in \emph{IEEE Conference on Computer Vision and Pattern Recognition (CVPR)}, 2018, pp. 7132--7141.

\bibitem{li2025coarse}
H.~Li, G.~Zhang, K.~Zhou, Y.~Cai, and G.~Yu, ``Coarse-to-fine: A dual-phase channel-adaptive method for wireless image transmission,'' \emph{IEEE Transactions on Wireless Communications}, vol.~25, pp. 2607--2623, 2026.

\bibitem{3gpp.38.214}
{3GPP}, ``{NR; Physical layer procedures for data},'' {3rd Generation Partnership Project (3GPP)}, Technical Specification (TS) 38.214, Mar. 2024, ver. 18.2.0.

\bibitem{3gpp.38.212}
------, ``{NR; Multiplexing and channel coding},'' {3rd Generation Partnership Project (3GPP)}, Technical Specification (TS) 38.212, Mar. 2024, ver. 18.2.0.

\bibitem{van2017neural}
A.~van~den Oord, O.~Vinyals, and K.~Kavukcuoglu, ``Neural discrete representation learning,'' \emph{Advances in Neural Information Processing Systems (NeurIPS)}, vol.~30, pp. 6306--6315, 2017.

\bibitem{yu2018slimmable}
J.~Yu, L.~Yang, N.~Xu, J.~Yang, and T.~Huang, ``Slimmable neural networks,'' \emph{arXiv preprint arXiv:1812.08928}, 2018.

\bibitem{tolstikhin2021mlp}
I.~O. Tolstikhin, N.~Houlsby, A.~Kolesnikov, L.~Beyer, X.~Zhai, T.~Unterthiner, J.~Yung, A.~Steiner, D.~Keysers, J.~Uszkoreit, M.~Lucic, and A.~Dosovitskiy, ``{MLP-Mixer}: An all-{MLP} architecture for vision,'' \emph{Advances in Neural Information Processing Systems (NeurIPS)}, vol.~34, pp. 24\,261--24\,272, 2021.

\bibitem{huang2017arbitrary}
X.~Huang and S.~Belongie, ``Arbitrary style transfer in real-time with adaptive instance normalization,'' in \emph{IEEE International Conference on Computer Vision (ICCV)}, 2017, pp. 1501--1510.

\bibitem{hu2022lora}
E.~J. Hu, Y.~Shen, P.~Wallis, Z.~Allen-Zhu, Y.~Li, S.~Wang, L.~Wang, and W.~Chen, ``{LoRA}: Low-rank adaptation of large language models,'' in \emph{International Conference on Learning Representations (ICLR)}, 2022.

\end{thebibliography}

\end{document}